\documentclass[twocolumn,twocolappendix]{aastex631}

\newcommand{\Cii}{C\,{\sc ii}}

\newcommand{\Civ}{C\,{\sc iv}}
\newcommand{\Siiv}{Si\,{\sc iv}}
\newcommand{\Oi}{O\,{\sc i}}
\newcommand{\Oiii}{[O\,{\sc iii}]}
\newcommand{\QSO}{QSO J0100$+$2802}

\usepackage{ulem}
\usepackage{comment}

\received{}
\revised{}
\accepted{}

\shorttitle{EIGER first results}
\shortauthors{Kashino et al.}

\graphicspath{{./}{./figs/}{./figures/}}

\begin{document}

\title{EIGER I. a large sample of {\Oiii}-emitting galaxies at $5.3 < z < 6.9$ and direct evidence for local reionization by galaxies}

\correspondingauthor{Daichi Kashino}
\email{kashino.daichi.v7@f.mail.nagoya-u.ac.jp}\tabletypesize{\footnotesize}

\author[0000-0001-9044-1747]{Daichi Kashino}
\affiliation{Institute for Advanced Research, Nagoya University, Nagoya 464-8601, Japan}
\affiliation{Department of Physics, Graduate School of Science, Nagoya University, Nagoya 464-8602, Japan}

\author[0000-0002-6423-3597]{Simon J.~Lilly}
\affiliation{Department of Physics, ETH Z{\"u}rich, Wolfgang-Pauli-Strasse 27, Z{\"u}rich, 8093, Switzerland}

\author[0000-0003-2871-127X]{Jorryt Matthee}
\affiliation{Department of Physics, ETH Z{\"u}rich, Wolfgang-Pauli-Strasse 27, Z{\"u}rich, 8093, Switzerland}

\author[0000-0003-2895-6218]{Anna-Christina Eilers}
\affiliation{MIT Kavli Institute for Astrophysics and Space Research, 77 Massachusetts Avenue, Cambridge, 02139, Massachusetts, USA}

\author[0000-0003-0417-385X]{Ruari Mackenzie}
\affiliation{Department of Physics, ETH Z{\"u}rich, Wolfgang-Pauli-Strasse 27, Z{\"u}rich, 8093, Switzerland}

\author[0000-0002-3120-7173]{Rongmon Bordoloi}
\affiliation{Department of Physics, North Carolina State University, Raleigh, 27695, North Carolina, USA}

\author[0000-0003-3769-9559]{Robert A.~Simcoe}
\affiliation{MIT Kavli Institute for Astrophysics and Space Research, 77 Massachusetts Avenue, Cambridge, 02139, Massachusetts, USA}

\begin{abstract}
 


We present a first sample of 117 {\Oiii}$\lambda\lambda$4960,5008-selected star-forming galaxies at $5.33 < z < 6.93$ detected in \textit{JWST}/NIRCam 3.5$\mu$m slitless spectroscopy of a $6.5\arcmin \times 3.4\arcmin$ field centered on the hyperluminous quasar SDSS J0100$+$2802, obtained as part of the EIGER (Emission-line galaxies and Intergalactic Gas in the Epoch of Reionization) survey.  Three prominent galaxy overdensities are observed, one of them at the redshift of the quasar.  Galaxies are found within 200~pkpc and 105~$\mathrm{km~s^{-1}}$ of four known metal absorption-line systems in this redshift range.  We focus on the role of the galaxies in ionizing the surrounding intergalactic medium (IGM) during the later stages of cosmic reionization and construct the mean Ly$\alpha$ and Ly$\beta$ transmission as a function of distance from the galaxies.  
At the lowest redshifts in our study, $5.3 < z < 5.7$, the IGM transmission rises monotonically with distance from the galaxies.  This is as expected when galaxies reside at peaks in the overdensity field of an IGM that is ionized by more or less uniform ionizing background, and has been seen at lower redshifts.
In contrast, at $5.7 < z < 6.14$, the transmission of both Ly$\alpha$ and Ly$\beta$ first increases with distance, but then peaks at a distance of 5~cMpc before declining. This peak in transmission is qualitatively similar to that seen (albeit at smaller distances and higher redshifts) in the \textsc{THESAN} simulations.  Finally, in the region  $6.15 < z < 6.26$ where the additional ionizing radiation from the quasar dominates, the monotonic increase in transmission with distance is re-established.  This result is interpreted to represent evidence that the transmission of the IGM at $z \sim 5.9$ towards J0100$+$2802 results from the ``local'' ionizing radiation of galaxies that dominates over the much reduced cosmic background.

\end{abstract}



\section{Introduction}
\label{sec:intro}

Understanding cosmic reionization is a central question in cosmology and galaxy evolution, and was a primary scientific motivation for the development of \textit{JWST}. 
Observational evidence from the analysis of the Gunn-Peterson troughs in quasar spectra \citep[e.g.,][]{2006AJ....132..117F,2015MNRAS.447..499M,2018MNRAS.479.1055B,2018ApJ...864...53E} as well as analysis of the Thomson optical depth for the cosmic microwave background radiation \citep{2020A&A...641A...6P} have suggested that the Epoch of Reionization (EoR) extended down to $z \sim 6$, with a tail extending even later to $z\sim5.5$ \citep{2017MNRAS.465.4838G}.  
This is also supported by the reports of declines in the Ly$\alpha$ equivalent widths of high-$z$ galaxies \citep[e.g.,][]{2012ApJ...744..179S,2014ApJ...794....5T,2018ApJ...856....2M} and in the Ly$\alpha$ luminosity function at $z \gtrsim 6$ \citep{2014ApJ...797...16K}.
In parallel, surveys of galaxies in the EoR, complemented by studies of their analogues at lower redshifts, have suggested that the bulk of the ionizing photon budget that is required to complete reionization could plausibly be coming from young star-forming galaxies, with reasonable ionizing photon escape fractions \citep[e.g.,][]{2015ApJ...802L..19R,2016MNRAS.461.3683I,2018MNRAS.478.4851I,2020ApJ...889..161N,2021ApJ...917L..37C,2022MNRAS.512.5960M}.    
However, direct evidence for a connection between galaxies and IGM opacity fluctuations in the same volume has not been presented.

To date, there have been some limited observations of the physical connections between the IGM and galaxies during the EoR on different physical scales.  Wide-field searches for Lyman-$\alpha$-emitters (LAEs) and Lyman-break galaxies (LBGs) along the sightlines towards $z\gtrsim6$ quasars have suggested an \emph{inverse} correlation at $z\sim5.7$ between the effective IGM opacity and the galaxy density, when both are averaged over very large scales $\gtrsim 50~\mathrm{cMpc}$, suggesting that reionization proceeds more slowly in underdense regions \citep{2018ApJ...863...92B,2020ApJ...888....6K,2021ApJ...923...87C,2022MNRAS.515.5914I}.
On smaller scales, \citet{2020MNRAS.494.1560M} measured the cross-correlation between Ly$\alpha$ transmission spectra of high-$z$ quasars and surrounding LAEs and LBGs with spectroscopic redshifts, and reported a possible excess transmission at $\sim10\textrm{--}60~\mathrm{cMpc}$ away from the galaxies, likely reflecting the effects of ionizing radiation from these galaxies (see also \citealt{2018MNRAS.479...43K}).  Similarly, excess Ly$\alpha$ transmission was also reported associated with {\Civ} absorbers, implying that it too reflects local ionization by the galaxies hosting the absorbers \citep{2019MNRAS.483...19M}.

Meanwhile, state-of-the-art radiation hydrodynamic simulations, coupled with realistic galaxy formation models, have become more reliable in predicting the connections between galaxies and IGM opacity, including the excess transmission near galaxies \citep{2022MNRAS.511.4005K,2022MNRAS.512.3243S,2022MNRAS.512.4909G}, following theoretical investigations using hydrodynamic radiative transport computations \citep{2014ApJ...793...29G,2017ApJ...841...26G,2019MNRAS.485L..24K,2019ApJ...876...31G,2020MNRAS.491.1736K}.
However, direct observational constraints of the local ionizing effects of galaxies have remained quite sparse, limiting our understanding of the relative contributions from different galaxy populations \citep[e.g.,][]{2020ApJ...892..109N} and the impact of individual galaxies on the surrounding IGM.

A different probe of diffuse gas during reionization is offered by the metal absorption systems that are seen in high-resolution spectra of high-redshift quasars.  
The abundance of metals and their ionization state must reflect a combination of the early chemical enrichment of the circumgalactic mediun (CGM), and the local ionizing radiation from the host galaxies of the absorbers \citep{2017ARA&A..55..389T}.  Particularly, as we approach the EoR, where the Ly$\alpha$ absorption saturates due to higher neutral fractions, metal absorption lines become the only discrete observational probe of the ambient baryonic matter.  Observations of these metal absorption lines have suggested that the incidence of low-ionization metal absorption systems, represented by species like {\Cii} and {\Oi}, is roughly constant, or even increases, towards the EoR, whereas that of high-ionization species, like {\Civ}, appears to drop quickly beyond $z\approx5$ \citep[e.g.,][]{2006ApJ...653..977S,2011ApJ...743...21S,2013ApJ...764....9M,2013MNRAS.435.1198D,2017ApJ...850..188C,2018MNRAS.481.4940C,2019ApJ...882...77C,2019ApJ...883..163B}. 
This observed difference in behavior is of great interest for understanding changes in the CGM that may be linked to the global reionization process.  Currently, however, the observational identification of the galaxies associated with the metal absorption systems at these redshifts is very limited \citep{2021NatAs...5.1110W,2021MNRAS.502.2645D}, because of the general difficulty of measuring spectroscopic redshifts of such high-$z$ faint galaxies.

Another aspect of galaxy evolution in the early universe involves the growth of the supermassive black holes (SMBHs) that power the very luminous quasars seen at these redshifts.   These earliest quasars provide strong constraints on SMBH formation scenarios, appearing to require super/hyper-Eddington mass accretion, direct collapse of a high-mass seed, or otherwise high duty cycles \citep[e.g.,][]{2011Natur.474..616M,2018Natur.553..473B,2020ApJ...897L..14Y}.     
Unfortunately, estimates of the masses of SMBH at these redshifts have
relied on rest-frame UV lines like Mg\,{\sc ii} and {\Civ}, which must be
indirectly tied back to the H$\beta$ line.
With \textit{JWST}, however, it has become possible to use the width of the rest-frame optical H$\beta$ line that provides, coupled with the luminosity of the quasar, the most robust estimates of the SMBH that can be directly compared to the local calibrations of SMBH mass from reverberation mapping \citep{2013ApJ...767..149B}.   
In parallel with the more reliable estimates of SMBH mass, observing the larger-scale environment and the nature of the host galaxies of these earliest quasars are of particular interest in the context of the structure formation in the $\Lambda$CDM paradigm and (co-)evolution of galaxies and SMBHs.

Motivated by these multi-faceted science topics, we initiated the survey project EIGER (\emph{Emission-line Galaxies and Intergalactic Gas in the Epoch of Reioniation}).  This is built around a 116 hour \textit{JWST} program of Guaranteed Time Observation (GTO)  (PI: Lilly; PID 1243), using deep wide-field slitless spectroscopy (WFSS) and broadband direct imaging observations using the Near InfraRed Camera (NIRCam) on \textit{JWST}. 
The key goal of this program is to examine the direct connections between galaxies and the state of gas in the circumgalactic and intergalactic environments towards the end of the EoR, by carrying out deep and complete surveys for \Oiii$\lambda$5008-selected emission-line galaxies in the fields of high-redshift luminous quasars for which we have also obtained extremely deep high-resolution spectra.

This Paper I is based on the first EIGER quasar field to have been observed (of six planned). We present an overview of the observational design of EIGER and our first sample of \Oiii$\lambda\lambda$4960,5008-selected galaxies at $5.3<z<6.9$ in this single field.  We focus in this paper on early results from the cross-correlation analysis between the galaxy sample and the transmission of the IGM. The detailed characterization of the {\Oiii}-emitting galaxies is presented in \citet[][hereafter Paper II]{EIGER_Paper2}. Other aspects of the data will be exploited in a series of companion papers.

The paper is organized as follows.
In \S \ref{sec:survey}, we provide the survey overview, and describe the \textit{JWST}/NIRCam observations, data reduction, and the identification of emission-line galaxies.  The ground-based spectroscopy of the target quasar is described in \S \ref{sec:qso_spectroscopy}.   \S \ref{sec:galaxy_sample_results} then presents the galaxy sample and a first look at the distribution of galaxies along the line of sight, including the identification of a number of strong overdensities, and also galaxies associated with discrete metal absorption systems. We then present a more detailed analysis of the transmission of Ly$\alpha$ and Ly$\beta$ in the IGM as a function of distance around the galaxies in \S \ref{sec:Lyaforest}.  \S \ref{sec:summary} summarizes our results.  Throughout the paper, we adopt a flat $\Lambda$CDM cosmology with $\Omega_\mathrm{\Lambda}=0.69$, $\Omega_\mathrm{M}=0.31$ and  $H_0=67.7~\mathrm{km~s^{-1}~Mpc^{-1}}$ \citep{2020A&A...641A...6P}.  All magnitudes are quoted in the AB system.

\section{The EIGER survey}
\label{sec:survey}

\subsection{Survey design}

The \textit{JWST} GTO program 1243, the EIGER survey, was primarily designed to explore the evolution of the IGM and of circumgalactic environments at the tail-end of cosmic reionization, and thereby to try to better understand the reionization process.  The 116 hr \textit{JWST} observing program consists of identical mosaics of simultaneous NIRCam WFSS and imaging observations \citep{2017JATIS...3c5001G} in six widely separated fields, each centered on a bright quasar at $5.98 < z < 7.08$.  
The program is complemented by $100+$ hours of ground-based deep high-resolution spectroscopy of all these quasars and, depending on the field, by additional deep \textit{Hubble Space Telescope} (\textit{HST}) imaging data, optical wide-field integral-field spectroscopic campaigns with VLT/MUSE and Keck/KCWI, and ALMA observations.

As a \textit{JWST} GTO program, the key features of the EIGER program were defined about five years ago. It became clear at that time that WFSS has a number of attractions for a study of this sort.  First and foremost, spectra are obtained for all objects in the field of view, completely avoiding the need for the constrained pre-selection of targets that is inherent in the operation of multi-slit spectrographs such as NIRSpec (as well as any practical difficulties associated with slits, including their alignment with targets and possibly poorly controlled slit losses).  This 100{\%} coverage of WFSS ensures that a complete census of galaxies in the field (above some straightforward flux criteria) is obtained.  In addition, although deep WFSS inevitably suffers from problems of contamination and over-lapping spectra, because of the high surface number density of faint sources in the sky, the relatively high spectral resolution of the NIRCam grisms suggested that it would be relatively straightforward to subsequently process the dispersed WFSS images so as to cleanly remove the continuum emission of all sources, leaving only the signals from emission lines.

The main difficulty of WFSS for a study of this kind concerns the identification of the source responsible for any given emission line, since there is an intrinsic degeneracy between spatial position and wavelength for any emission line detected in the dispersed image, plus the usual difficulties of line identification for single emission lines.  This latter issue motivated our primary focus on the {\Oiii}$\lambda\lambda$4960,5008 doublet, while allowing for the possibility of ancillary science using other lines at lower redshift.  As rest-frame visible emission it is relatively unaffected by dust reddening.  At the redshifts of interest, $5 < z < 7$, this emission is redshifted to 3.5 $\mu$m where the \textit{JWST} background is lowest.  Furthermore, previous photometric studies had suggested that these lines would be very strong, with high equivalent widths \citep[e.g.,][]{2013ApJ...763..129S,2014ApJ...784...58S}.

Most crucially, the fact that this is a well-defined doublet with a fixed line ratio, and often appears as a triplet with the nearby H$\beta$ line, suggested that {\Oiii}$\lambda\lambda$5008,~4960 emission could be easily isolated in the high-resolution ($R\sim1500$) NIRCam WFSS dispersed images.
Furthermore, the observed $\Delta \lambda$ spacing between these two lines on the detector would already give an indication of the redshift and thus wavelength, significantly constraining the spatial positions of the sources responsible.  Of course, for a given line flux, WFSS is always less sensitive than slit spectroscopy, because of the elevated background in WFSS images, but the extraordinary sensitivity of \textit{JWST} relative to any previous facility made this an acceptable trade, especially since, except at low spectral resolution, dispersed slit spectroscopy with NIRSpec is detector-noise limited.

NIRCam WFSS observations with the F356W ($\approx 3.1\textrm{--}4.0~\mathrm{\mu m})$ filter in the Long Wavelength (LW) channel therefore allow isolation of {\Oiii}$\lambda\lambda$5008,~4960 at $5.3 < z < 7.0$, the target redshift for the EIGER survey of galaxies at the end of the EoR.  A large number of other emission lines at generally lower redshift are also detected in the WFSS emission-line image.  Scientific utilization of these other emission lines is of course possible, but requires more careful treatment of the identification problem and will be addressed in later papers.

Relatively short direct images are also mandatory in the same F356W filter to allow precise positional identification of sources.  The two channels of NIRCam allow deep direct images to be obtained in the F115W and F200W filters in the Short Wavelength (SW) channel during the WFSS and direct imaging in F356W in the Long Wavelength (LW) channel.  
The two SW filters were chosen to allow the characterization of the spectral energy distribution (SED) of galaxies at $z\sim6$ by sampling the rest-frame UV spectrum in the two SW bands and, together with the F356W photometry, across the Balmer/4000{\AA} break band (although the F356W fluxes by construction contain multiple strong emission lines).

As described in detail below, a mosaic pattern of four offset field centers (each representing a separate JWST ``Visit'') was chosen so as to cover a contiguous reasonably wide survey area at more or less uniform depth, together with a central deep overlap region that was centered on each of the quasars.   The full survey area of $6.5 \times 3.4~\mathrm{arcmin^2}$ around each quasar enables the characterization of the large-scale galaxy distribution on scales of up to $\sim10~\mathrm{cMpc}$ around the sightline, enabling the cross-correlation of galaxies and the Ly$\alpha$ and Ly$\beta$ transmission of the IGM to be robustly examined on these scales with a complete galaxy sample (avoiding the danger of missing galaxies).  The central deeper overlap area of $\approx 40 \times 40~{\rm arcsec}^2$ allows the full depth to be achieved close into the sightline, enabling the nature of the quasars, their host galaxies, and their surrounding environments to be examined, and to carry out high-completeness searches for galaxies associated with metal absorption systems in the quasar spectra.   
The overall exposure time was set by the maximum depth that we required close to the quasar.  The size of the positional offsets for the four centers in the mosaic pattern therefore at some level represents a trade-off between reducing the area of this central, maximally exposed, region and increasing the total area observed at 25{\%} exposure level.  In practice the offset, at least in the $x$-direction, is largely determined by the gap between the two modules of the NIRCam instrument.

\subsection{Target quasar selection}

The six target quasars (Table \ref{tb:target_quasars}) were selected to address a range of complementary science goals.  All six had very deep high-resolution spectra available from optical to NIR from ground-based observations with VLT/X-Shooter, Magellan/FIRE, and Keck/HIRES.  The quasars were chosen on the basis of what was known about the properties of the Ly$\alpha$ forest and/or Gunn-Peterson trough, the existence and ionization state of the known metal absorption systems along the line of sight, and the characteristics of the quasars themselves.

In the following, we describe the first EIGER observations that have been successfully completed in the field of  SDSS J010013.02+280225.8 (hereafter J0100+2802) at $z=6.327$ \citep{2015Natur.518..512W,2020ApJ...904..130V} during August 2022.  It is anticipated that an almost identical set of data will be obtained on the other five quasars during the first year of \textit{JWST} science observations.

\begin{deluxetable}{ccc}
\tablecaption{Target quasars \label{tb:target_quasars}}
\tablehead{
    \colhead{Identifier}&
    \colhead{$z_\mathrm{QSO}$}&
    \colhead{References}\tablenotemark{a}}
\startdata
 SDSS J010013.02+280225.8 &  6.327 & 1, 2 \\
 ULAS J014837.63+060020.0 &  5.98 & 3 \\
 SDSS J103027.09+052455.0 &  6.280 & 4 \\
 PSO J159.2257-02.5438    &  6.381 & 5 \\
 ULAS J112001.48+064124.3 &  7.085 & 6, 7 \\
 SDSS J114816.64+525150.3 &  6.475 & 8 \\
\enddata
\tablenotetext{a}{References: 1--\citet{2015Natur.518..512W}; 2--\citet{2020ApJ...904..130V}; 3--\citet{2015AJ....149..188J}; 4--\citet{2001AJ....122.2833F}; 5--\citet{2016ApJS..227...11B}, 6--\citet{2011Natur.474..616M}; 7--\citet{2020ApJ...904..130V}; 8--\citet{2015ApJS..219...12A}}
\end{deluxetable}

\subsection{The NIRCam observations}
\label{sec:observations}

\begin{figure*}[t]
\centering
\includegraphics[width=7.0in]{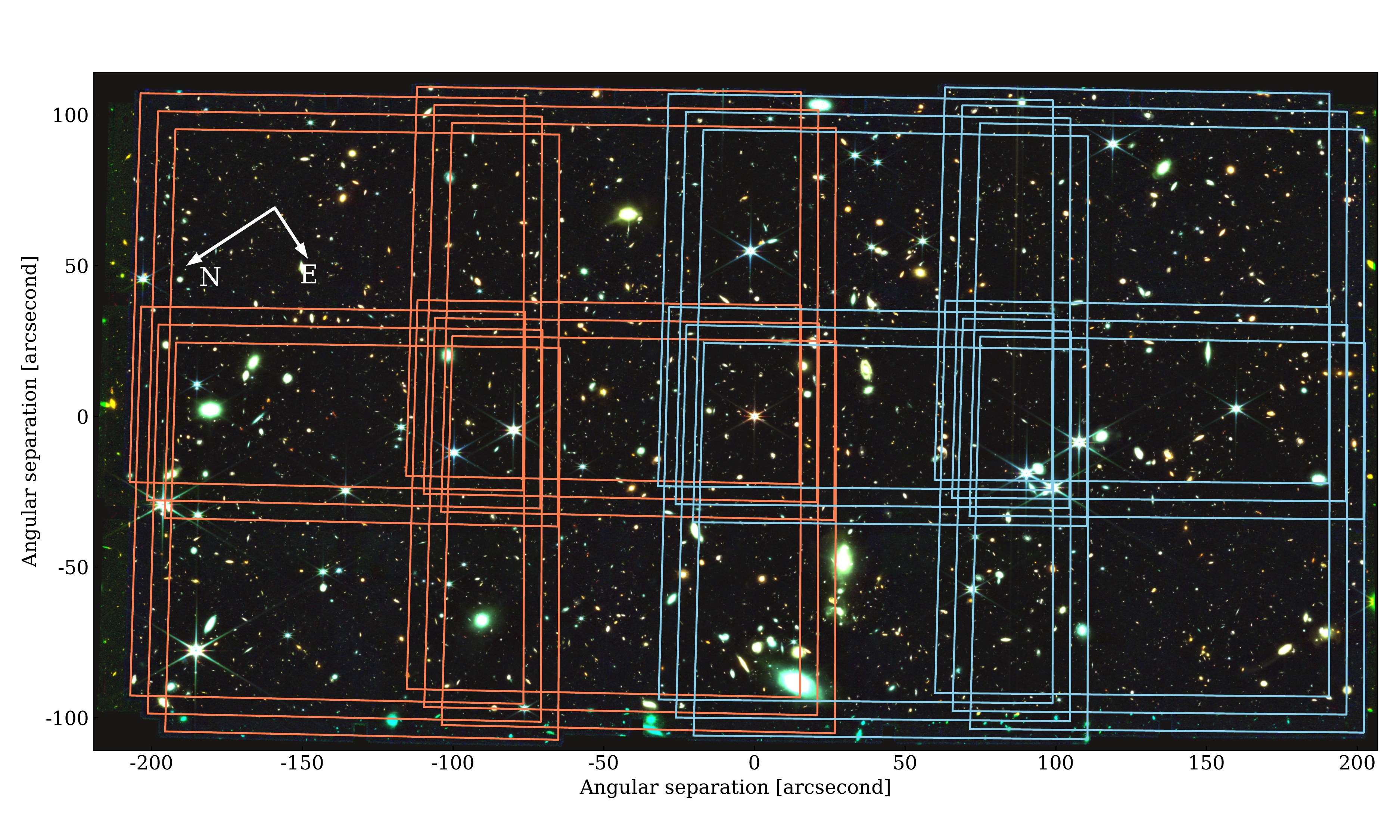}
\caption{
\textit{JWST} NIRCam observations of the field of the quasar J0100$+$2802.  The background is the coadded pseudo-color image in the F115W (blue), F200W (green) and F356W (red) bands.  The axes values are the angular separations from the position of the quasar.  The detector fields-of-view of the individual WFSS exposures in our mosaic observations are shown by the large red and blue squares, respectively, for the NIRCam module A and B detectors.  
\label{fig:main_map}}
\end{figure*}

The NIRCam imaging and WFSS observations of the field of QSO J0100$+$2802 were carried out during August 22--24, 2022.

NIRCam has two modules (Modules A and B) each with a $2.2\arcmin \times 2.2\arcmin$ field of view separated by $44\arcsec$. 
The EIGER observations consist of a mosaic of four otherwise identical ``Visits'' that form a $2 \times 2$ overlapping mosaic pattern.  
This gives a contiguous survey field of $\approx 6.5\arcmin \times 3.4\arcmin$ centered on the target quasar and  a central $\approx40\arcsec \times 40\arcsec$ region with fourfold longer exposure time.
The position angle of the mosaic (defined by the telescope V3 axis) for the observations of QSO J0100$+$2802 was 236 degrees.  This is usually arbitrary but was constrained for some of the quasar fields to ensure that the central region around the quasar in each grism image was not contaminated by spectra from other known  brighter sources in the field. The mosaic pattern in the {\QSO} field is shown in Figure \ref{fig:main_map}.

The WFSS observations use the LW channel and a combination of the grism R and F356W filters.  This combination yields $3.1\textrm{--}4.0~\mathrm{\mu m}$ spectra dispersed along the detector rows.  The spectral resolution for a point source is $R\sim1500$
at around $3.5~\mathrm{\mu m}$, and the dispersion is $\approx 1~\mathrm{nm~pixel^{-1}}$.
The two modules A and B, with the grism R, disperse the spectra in opposite directions, i.e., the wavelength increases towards the $+x$ and $-x$ directions in the detector coordinates, respectively.   In both modules the zero deviation wavelength is 3.95~$\mu$m, close to the long wavelength end of the F356W filter response curve.

The adopted mosaic configuration thus yields two reversed dispersion spectra for all sources within a central vertical strip of width $\approx 60\arcsec$ that is covered by both modules.  Furthermore, two independent but not-reversed spectra are obtained for all sources within two off-center vertical strips of width also $\approx 60\arcsec$ and a central horizontal strip of width $\approx 70\arcsec$.   The horizontal and center vertical strips intersect in the central square region centered on the quasar.

Together with the longer exposure time, the availability of the two reversed grism spectra helps to disambiguate the identification of source objects in the direct images, as well as the wavelengths of any emission lines, and to  remove contaminating lines from other sources.  When only one dispersion direction is available (which is the case over most of the extended survey field) the identification of the correct source in the direct image (to produce the correct wavelength solution) requires some care. However, extensive simulations before \textit{JWST} launch indicated that this could be achieved with high confidence, and this has proved to be the case in practice.

Likewise, since the primary science goal was the detection of a very recognisable spectral feature, the \Oiii$\lambda\lambda$4960, 5008 doublet (or the triplet including H$\beta$) the difficulties produced by overlapping spectra (from interloping continuum source or line emission) were minimized.  
As will be described below, the relatively high spectral resolution of the NIRCam WFSS means that contaminating continuum sources with slowly varying SEDs can be removed effectively by suitable filtering of the data.

The effective field-of-view for the dispersed images depends upon the observed wavelength and therefore the redshift of each source.  The dispersed spectra with grism R through the F356W filter are $\sim940$ pixels long, about half the width of the detector (2048~pixels), and extend across from the undeviated wavelength of 3.95~$\mu$m.  Therefore, in Module A the dispersed spectrum extends across $-850$ to $+90$ pixels relative to the undeviated position of the source, or $+850$ to $-90$ pixel in Module B.   

Given the LW detector size of $2048^2$~pixels, the spectra of sources within $\sim 850$~pixels from one of the vertical edges (at $x=1$ for Module A and at $x=2048$ for Module B) extend beyond the detector coverage, losing (in both modules) some length of the spectrum at the short wavelength end.   When the redshift is based on a single spectral feature, this effectively reduces the survey area at the low redshift end of the redshift range.
The $2\times2$ mosaic pattern used in the EIGER survey means that the full spectra are in fact obtained for half the affected sources in another pointing.  In the combined data set, full spectral coverage is obtained for a region of width $\approx 4.6\arcmin$.   Provided the emission line is bright enough to be detected, the effective survey area therefore increases from $\approx 17~\mathrm{arcmin}^2$ at $\lambda=3.1~\mathrm{\mu m}$ up to the maximum $\approx 25~\mathrm{arcmin}^2$ at $\lambda=3.95~\mathrm{\mu m}$, corresponding to redshifts of \Oiii$\lambda$5008 from $z=5.3$ to $6.9$.

During the WFSS observations, simultaneous direct (non-dispersed) imaging was obtained using the SW channel, first with the F115W filter and then with F200W.  After the WFSS observations were completed, additional direct imaging was then obtained in the F356W filter in the LW channel, with further exposure time obtained in F200W in the SW channel. This last set of direct images include two displaced field centers to ensure that direct images are obtained for all sources whose light can be dispersed by the grism onto the detector.

In summary, each visit has an identical sequence that consists of 1) F356W WFSS + F115W imaging, 2) (identical) F356W WFSS + F200W imaging, and 3) direct imaging in F356W + F200W. 
For both the WFSS and SW direct imaging, we employed a combination of the 3-point INTRAMODULEX primary and 4-point subpixel dithering, yielding a total of 12 exposures per visit and module in each of F115W and F200W and 24 WFSS images. 
Co-adding these multiple exposures, the total nominal exposure time in a single visit is 4380 seconds per filter for the SW imaging and 8760 seconds for the WFSS.

Given the small-scale dithering and larger-scale mosaic pattern, the total exposure time varies across the field, ranging from 1.5ks (kilo-seconds; for the extreme edge regions sampled by only a single primary dither) up to 13ks for SW imaging and from 2.9ks upto 35ks for WFSS.   The F356W direct and two ``out-of-field'' images have an exposure time of 526 seconds each, yielding a total exposure time of 1578 seconds in the majority of the coverage these three overlap, and a maximum 6.3ks in the overlap regions of the four visits.

\subsection{Imaging data reduction and photometry}
\label{sec:imaging_reduction}

The \textit{JWST}/NIRCam imaging data were reduced using the \texttt{jwst} standard calibration pipeline (version 1.8.2\footnote{\url{https://github.com/spacetelescope/jwst}}), adopting the Calibration Reference Data System (CRDS) context \texttt{jwst\_0988.pmap (PUB)} including updated zero-points based on in-flight data \citep{2022RNAAS...6..191B}.
All the individual broadband raw images were first processed for detector-level corrections using the \texttt{Detector1} module of the pipeline.  Then, the individual products (count-rate images, i.e., \texttt{\_rate.fits} files) were calibrated through \texttt{Image2}, where flat-fielding and the flux-calibrations are applied to convert the data from units of count rate to surface brightness (or flux density).  Circular defects, known as ``snowballs'', that arise from strong cosmic ray hits were masked following \citet{2022ApJ...938L..14M}.  The individual calibrated images (\texttt{\_cal.fits} files) were then resampled and coadded onto the common pixel grid with a pixel resolution of $0.03\arcsec~\mathrm{pixel^{-1}}$ through the \texttt{Image3} processing.  The astrometry was calibrated to align stars from the Gaia Data Release 2 catalog \citep{2018A&A...616A...1G}.

The individual frames were further processed to subtract artificial scattered light, known as ``wisps'', $1/f$ noise (correlated horizontal and vertical noise; e.g. \citealt{2020AJ....160..231S}), and residual sky background, using a source mask based on a coadded image, before obtaining the \textit{final} coadded images.  
The wisps were removed using the templates that were constructed by median-combining all exposures in the detector coordinates in a given filter and detector element. These templates were only applied to the short wavelength filters, as the LW channel appears unaffected.  This worked well since our program conducted mosaic observation with large offsets in the sky.  The $1/f$ noise is excluded by a sigma-clipped median subtraction in each 1/4 detector patch of a row that corresponds to one amplifier, then likewise a subtraction per column, and finally per amplifier. The pseudo-color image constructed from the final coadded images in the F115W (blue), F200W (green), and F356W (red) bands is shown in Figure \ref{fig:main_map}.

Aperture-matched photometry was carried out using \texttt{SExtractor-2.25.0}\footnote{\url{https://sextractor.readthedocs.io/en/latest/index.html}} with the F356W imaging data as detection image.
First, we convolved the higher-resolution F115W and F200W images to match the point spread function (PSF) of the F356W image.
We derived the modeled PSFs for each of the filters using \texttt{webbpsf}\footnote{\url{https://webbpsf.readthedocs.io}} \citep{2014SPIE.9143E..3XP} and then created a matching kernel using a \texttt{SplitCosineBell} window (in \texttt{photutils}).
This methodology was tested extensively using \texttt{MIRaGe}\footnote{\url{https://mirage-data-simulator.readthedocs.io/en/latest/}} simulations.
Total magnitudes were measured with Kron apertures following \cite{2022ApJ...928...52F}. The Kron scaling factor was optimised to maximize signal-to-noise ratio (S/N) and we applied aperture corrections.  The uncertainty on the photometry was estimated using a model that fits the noise in random apertures in blank sky regions of the images as a function of aperture size and scaling for variations in the sensitivity.  The average 5$\sigma$ sensitivities are 27.8, 28.3, 28.1 in the F115W, F200W and F356W imaging data, respectively, reaching a maximum sensitivity 28.7, 29.2, 29.0 in the 10 \% deepest regions. 

\subsection{WFSS data reduction}
\label{sec:wfss_reduction}

\begin{figure*}[t]
\centering
\includegraphics[width=6.0in]{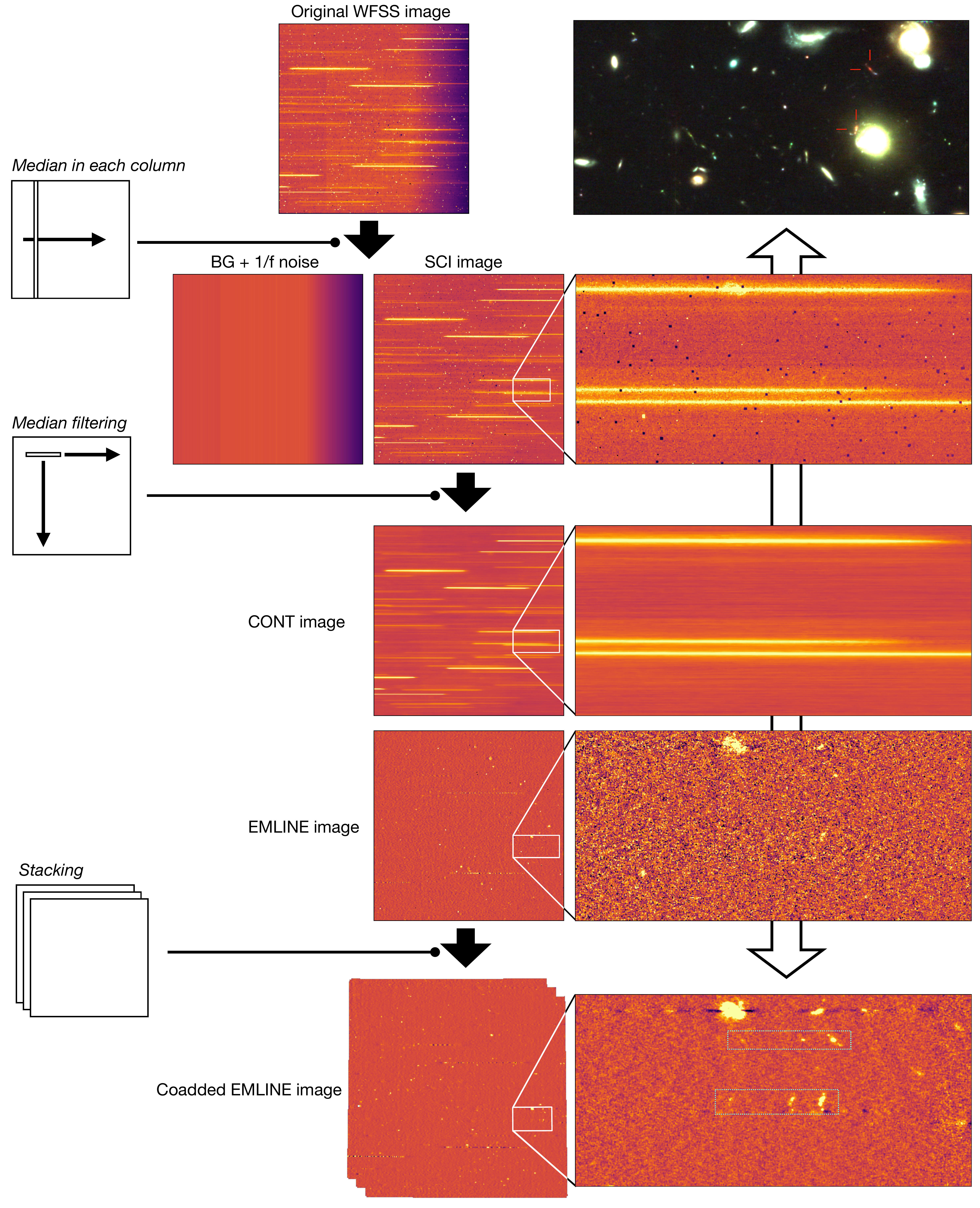}
\caption{
Processing of the WFSS data, after the \texttt{Detector1}, \texttt{assign\_wcs} and flat-fielding processes.  The WFSS images of individual exposures are first processed to subtract the background and $1/f$ noise.  The resulting SCI images, are then median-filtered to separate them into continuum (CONT) and emission line (EMLINE) images - see text for details.  These images are finally stacked using \texttt{Image3} to produce a single image for each visit and module.  The rightmost column shows a zoom-in of a small region of the WFSS images, with a direct image of the corresponding region in the top panel.  The rectangles (green dotted line) mark the {\Oiii} doublet and H$\beta$ lines that originate from the galaxies marked by the red reticles in the direct sky image. \label{fig:wfss_processing}}
\end{figure*}

\begin{figure}[t]
\centering
\includegraphics[width=3.5in]{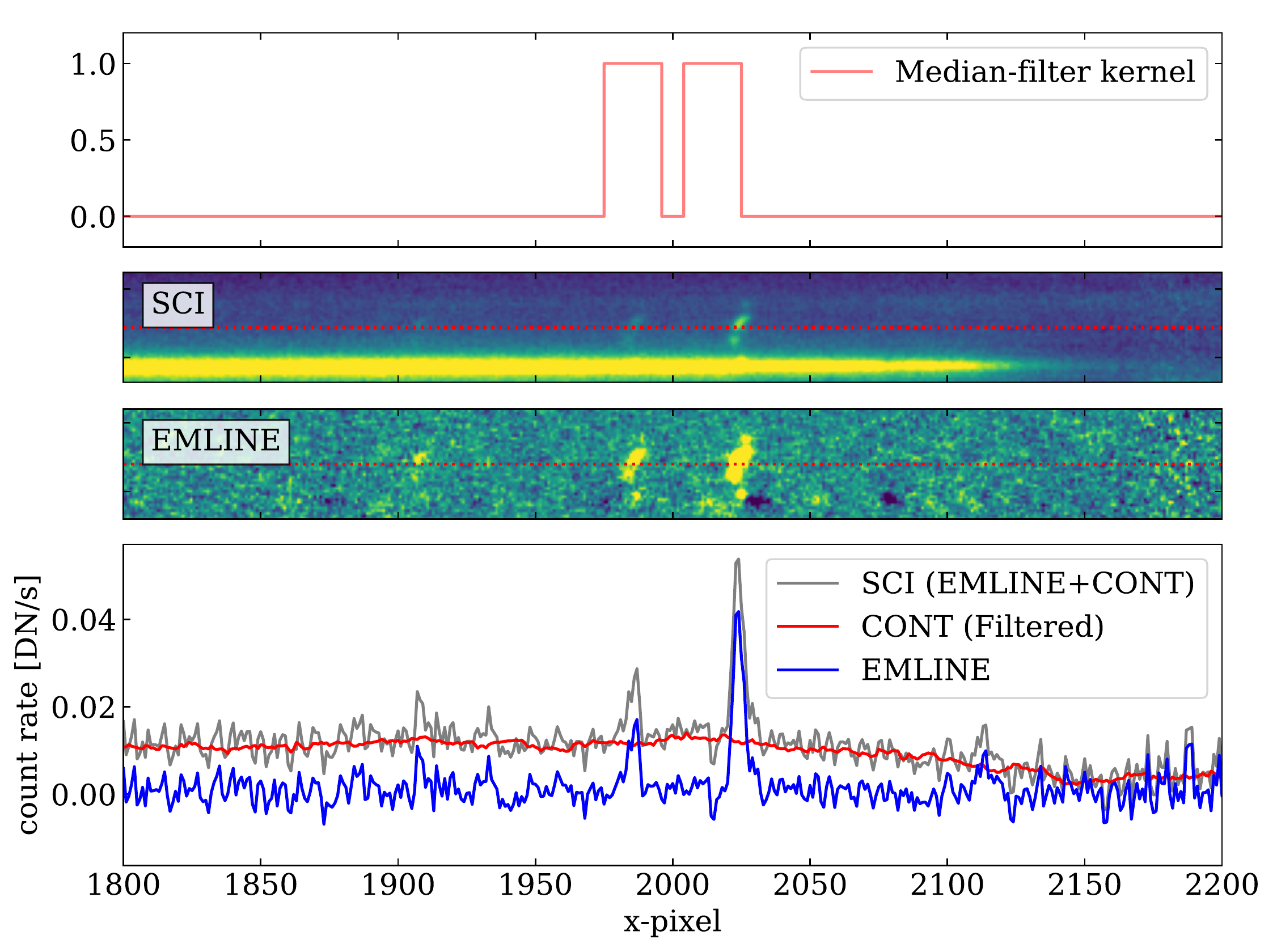}
\caption{Demonstration of the median filtering step to isolate emission lines and continuumThe top panel shows the kernel for median-filtering that has a full width 51~pixels and a central ``hole'' of 9 pixels.  The second and third panels show the zoom-in of a small region of the SCI and EMLINE grism images around the {\Oiii} doublet $+$ H$\beta$ lines.  The bottom panel shows 1D spectra, from the SCI (gray), CONT (red), and EMLINE (blue) frame, along the single ``row'' marked by the horizontal red dotted lines. 
All these panels have the same $x$-axis range across 400 pixels.  \label{fig:median_filtering}}
\end{figure}

The WFSS data were reduced with a combination of the \texttt{jwst} pipeline (version 1.7.0) and custom python-based processing steps.  Figure \ref{fig:wfss_processing} summarizes the procedure of the WFSS data processing.  The development of this method was motivated to allow for the robust detection of emission lines in the dispersed grism images.

We first processed individual grism images using \texttt{Detector1} for detector-level corrections, and used the \texttt{assign\_wcs} step from \texttt{Spec2} to obtain the WCS solution for each grism image.  We then processed the data using \texttt{Image2} for flat-fielding, and subsequently removed the large-scale sky background variations and $1/f$ noise by a straight median-subtraction in each detector column (orthogonal to the dispersion direction).  We refer to the output after this basic processing step as `SCI' image.

The next step is to isolate emission lines in the SCI images\footnote{The python programs for these steps will be made public upon publication.}. 
We applied an empirical median-filtering using a boxcar-like kernel along each detector row (i.e. parallel to the dispersion direction) to obtain filtered images that represent all the different continuum components that appear in the grism data.   This filtered continuum image may then be subtracted from the SCI image to yield, in principle, a pure emission line image.
The primary kernel for this median filtering had a length of 51 pixels with a central ``hole'' of 9 pixels ($\approx 770~\mathrm{km~s^{-1}}$) so as to avoid self-subtraction around emission lines.  Figure \ref{fig:median_filtering} presents this kernel shape and demonstrates that this median filtering generally worked very well in removing the continua.  This relatively long kernel, however, noticeably over-smooths the continua at each end of the spectrum, where the filter response curve produces large short-scale variations in the continuum level. This causes a spurious positive-negative pattern in the continuum subtracted images at each end of the spectrum.  To mitigate this end effect, we first identified the ends of continuum spectra by measuring the gradient of the brightness along the detector row, and adopted a shorter kernel of 21 pixels in these regions.  A composite continuum image was then produced from these two different-sized kernels with gradually changing fractions at the boundary to avoid discontinuity.  This step was actually repeated with a emission-line mask obtained from the coadded grism image as described below.
We note that the use of only the shorter kernel does not yield better results, particularly around the {\Oiii}+H$beta$ lines, as it is much more easily affected by emission lines with close separations than the longer kernel.
As shown in Paper II, this custom filtering could result in $<5\%$ level oversubtraction of the continuum around emission lines.

The emission-line images (hereafter `EMLINE') were then generated by subtracting these median-filtered continuum images (`CONT') from the SCI data.  These procedures were developed and tested extensively before launch using \texttt{MIRaGe} simulations based on the JAGUAR catalog \citep{2018ApJS..236...33W} and worked well on the real data.  The most noticeable residuals from the continuum subtraction procedure involved bright galaxies around $z \sim 0.5$ where the strong CO bandheads are not completely removed.  These residual features are easily recognized.

With the EMLINE images of individual exposures in hand, we then adopted two different approaches for image stacking and subsequent line identification.
In the first method, we used the \texttt{Image3} module of the \texttt{jwst} pipeline to coadd the set of 24 individual full-detector-size grism images (SCI, EMLINE, and CONT) per visit and module.  Stacking of the full-size images works only for images with small spatial offsets, up to the primary dither size, and the same dispersion direction.  The geometric relation between the position of a source and the location of the spectral trace on the detector has both spatial- and wavelength-dependent distortions.  But the {\it differential} distortion over the spatial scale of the dither pattern is small enough to be ignored allowing the co-adding of these images.

The coadded EMLINE images produced in this way enable us to directly detect all emission lines in the co-added grism data using \texttt{SExtractor}, without any knowledge of the sky locations of the source galaxies (see the bottom panels in Figure \ref{fig:wfss_processing}).

The second method extracts 2D spectra from the individual grism images for all sources detected in direct imaging (see \S \ref{sec:imaging_reduction}).  Despite the somewhat shorter exposure time of the F356W direct images, even pure emission line sources from the WFSS frame are essentially always detected in the direct image, since the backgrounds are essentially the same in the two. 
In spectral extraction, we used \texttt{grismconf}\footnote{\url{https://github.com/npirzkal/GRISMCONF}} to calculate a position ($x,y$) in a grism image for a given sky position and wavelength.  We adopted the NIRCam grism model (V4)\footnote{\url{https://github.com/npirzkal/GRISM_NIRCAM}}, based on commissioning observations.  
The spectral trace is not perfectly parallel to the detector row direction.  Therefore, the extracted 2D spectra were first rebinned column-by-column to be flat, and then resampled into a common linearly-spaced wavelength grid in steps of 9.75~{\AA}.  These procedures were executed so that the fluxes are conserved.  The data in units of count rate was then converted into flux density units ($\mathrm{erg~s^{-1}~cm^{-2}~{\AA}^{-1}}$) by using the total sensitivity curve as a function of wavelength.  The sensitivity curve is provided for each of Modules A and B in the grism model, and we slightly modified them by re-calibrating with commissioning data.
The average throughput is $\approx28\%$ lower in Module B than A.
Stacking was then performed for each source selected in the direct image using all available WFSS exposures, separately for each module: recall that the emission-line images, tracing the source shape in the sky, are flipped between the modules.  Emission lines could then be identified in these stacked 2D spectra (although the same emission line can appear in multiple stacks).

These two methods of the grism image stacking are linked to the two approaches for identification of emission lines and and the source galaxies, as described in the following subsection.

\subsection{Identification of \Oiii-emitting sources}
\label{sec:line_identification}

We developed two complementary methods of identifying \Oiii-emitting galaxies from the EMLINE frames, referred to hereafter as the ``backward'' and ``forward'' approaches.  These identifications were independently carried out by two members of our team (DK and JM, respectively), and then the candidate samples were reconciled to produce a final sample.

\begin{figure}[t]
\centering
\includegraphics[width=3.4in]{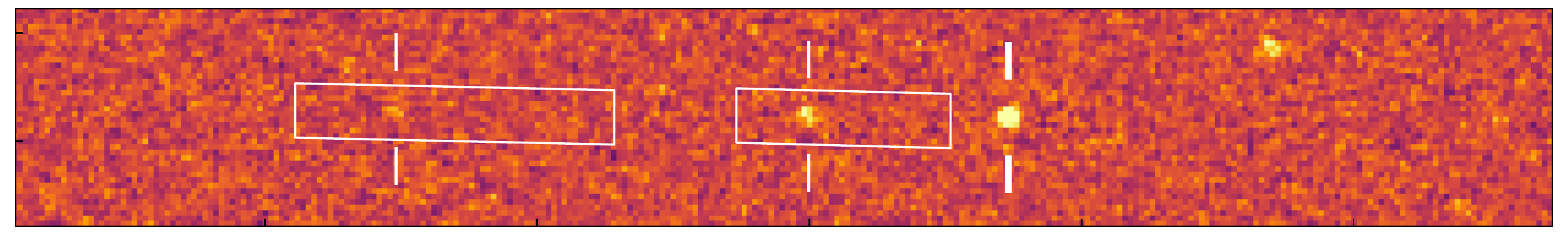}
\includegraphics[width=3.4in]{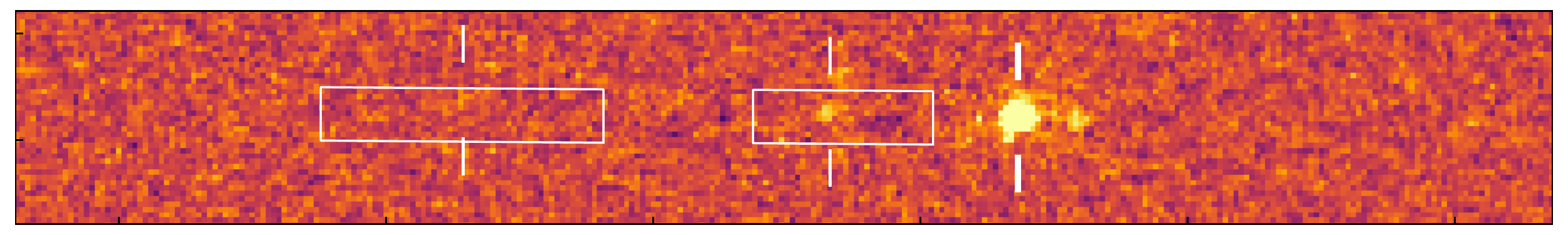}
\includegraphics[width=3.4in]{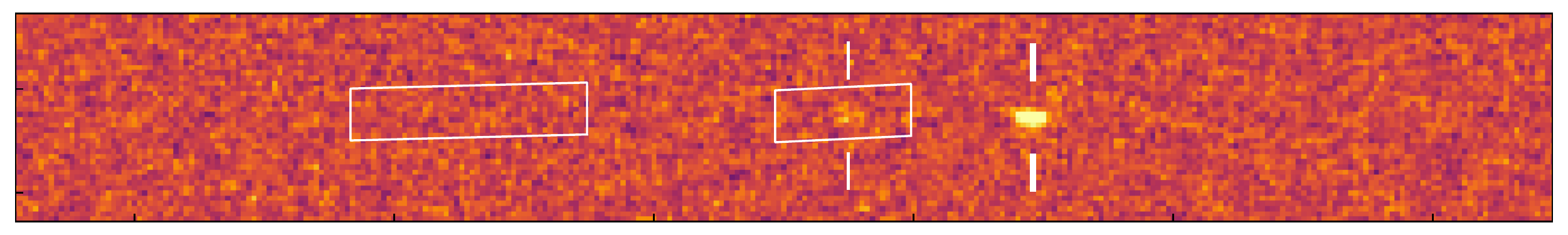}
\caption{
Three representative examples of the search for pairs or triplets of emission lines in the emission line images in the so-called ``backwards" approach. These are candidates for the {\Oiii} doublet and H$\beta$ .  The primary line detections, putatively {\Oiii}$\lambda$5008, are marked by thick reticles. The white boxes then indicate the search boxes for the secondary lines, i.e., {\Oiii}$\lambda$4960 and H$\beta$ which include the vertical uncertainty in position and the horizotal uncertainty in the redshift and resulting $\Delta \lambda$.  The left and right boundaries of these boxes therefore correspond, respectively, to the extreme cases that the primary {\Oiii}$\lambda$5008 is actually at $3.1~\mathrm{\mu m}$ and at $4.0~\mathrm{\mu m}$.   The secondary line detections found within these search boxes are marked by thin reticles.
All these three sources were indeed identified as secure{\Oiii} emitters.
\label{fig:line_search}}
\end{figure}

The backwards approach starts with the coadded EMLINE images per visit and module.  We first detected all emission-line signals using \texttt{SExtractor} and searched for pairs or triplets of detections which could conceivably be a combination of the {\Oiii}$\lambda\lambda$4960, 5008 doublet and/or H$\beta$.
In practice, we first selected all emission lines detected at $5\sigma$ or more as primary detection, where the S/N is determined by using the flux and associated errors in a fixed 5-pixel ($0.315\arcsec$) aperture.  
We then provisionally assigned this primary line to be {\Oiii}$\lambda$5008, and searched for the required secondary lines, i.e., {\Oiii}$\lambda$4960 and/or H$\beta$.  Because at this point the true source position is unknown, the putative {\Oiii}$\lambda$5008 line could lie anywhere in the possible $3.1\textrm{--}4.0~\mathrm{\mu m}$ window, i.e., anywhere in the $5.3 < z < 7.0$ redshift range, meaning that the other lines of the doublet/triplet could be displaced by a range of pixels.  The search boxes for the other putative components were therefore defined accordingly using the grism model.  Figure \ref{fig:line_search} shows examples of the detection of the primary lines and the search for the secondary lines.
Note that these procedures are based purely upon the emission lines and do not rely on any knowledge from direct imaging.  All pairs/triplets of lines that could conceivably be the desired {\Oiii}$\lambda$4960,5008 combination, with or without H$\beta$, were thereby recorded.  For each combination, an approximate redshift is derivable from the line separation in the grism image.

The next step is to examine the robustness of these line pairs/triplets and to identify the source galaxies of these emission lines.  The grism model enables us to inversely calculate the sky location (RA, DEC) for a given position in the grism image ($x,y$) for any and all wavelengths.  We therefore first derived a ``trace'' in the sky (i.e. a locus in the F356W direct image) as a function of wavelength between $3.1\textrm{--}4.0~\mathrm{\mu m}$ for each of the primary lines that had at least one possible secondary line detection of \Oiii$\lambda4960$ and/or H$\beta$, and then identified all sources which are located within $\approx 0.15\arcsec$ of this trace as candidates to be the source galaxy responsible for the emission lines in question.  
Finally, we then visually inspected possible combinations and determined which galaxy is the source of the emission lines, otherwise discarded as a non-{\Oiii} source.   A key requirement for all sources is that the implied redshift (fixed once the source position is determined) be consistent with the approximate redshift determined from the separation of the lines in the grism image.
Further judgements were made on the basis of the available information, including the spatial offset from the expected position, the flux ratio of the doublet lines, the morphological properties of the source in the direct and dispersed images, and the observed broadband color of the candidate sources.

These procedures for line detection and line-source identification were performed separately for each visit and each module.  The results of these visual inspections were then merged into a single combined catalog of {\Oiii}-emitter candidates by excluding duplicates and resolving conflicts in the overlap regions.

This ``backwards'' approach provides a sensible lower limit to the number of {\Oiii} emitters and has a potential advantage that it produces a purely line-flux limited set of emission-line sources including sources which may be ambiguous in the direct images due to its intrinsic faintness, blending with other bright sources, or any possible contamination.  On the other hand, the quality of the spectral images can not be optimal because of the stacking that can only be done per visit/module and the small but non-zero smearing of line signals due to the differential distortions, even between primary dithers ($\approx 6\arcsec$ in both $x$ and $y$ directions) within the same visit.

We therefore also attempted the inverse approach using the best-quality coadded images of the spectra as follows.
In this so-called ``forwards'' approach, we start with a catalog of broadband sources detected in the F356W image (see \S \ref{sec:imaging_reduction}).  We first picked up all $\sim20000$ sources brighter than $m_\mathrm{F356W} = 29~\mathrm{mag}$.  For each of these, we extracted 2D spectra from the grism (SCI, EMLINE, and CONT) images of the individual exposures, and then coadded them, as described in \S \ref{sec:wfss_reduction}.  We then ran \texttt{SExtractor} on each of these coadded EMLINE 2D spectral images to detect emission-line signals and selected those in which candidate line pairs/triplets were detected at $>3 \sigma$ close to the expected center of the spectral trace of the object and with a sensible separation in wavelength.

We conservatively selected all candidate pairs for the {\Oiii} doublet with a relative flux between a factor 1.5--6 (i.e., a factor 2 around the expected 2.98 ratio), a maximum 3 pixel ($0.19\arcsec$) difference in the spatial centroid (i.e,. in the column ($y$) direction) of the lines in the 2D spectrum, and with a doublet separation within 15{\AA} of the expected separation for the possible redshift solution.  Through visual inspection, we removed candidates where one of the lines was likely to be artifact, clearly came from a different galaxy, or was evidently a residual from the continuum filtering method, using the reversed dispersion images from both modules if available.  Then we verified the association between the lines and the broadband source primarily based on the spatial position and doublet separation and, in some cases, based on color and/or morphology.  In doing so, we carefully resolved the cases that a single emission-line pair is associated to multiple galaxies.  The main benefit of the forwards approach is that it can optimally combine data over multiple visits and modules and therefore yield fainter sensitivity than the backwards approach described above.

The two samples of {\Oiii}-emitter candidates, obtained from these two complementary approaches, were then carefully reconciled, with an iterative fine-tuning of the search parameter, so as to obtain the final catalog of {\Oiii} emitters.  A first comparison showed that the two samples agreed on $\sim 70\%$ of the final sample, with $\sim20\%$ of the fainter sources not initially identified in the backwards approach, and $\sim10\%$ missed by the forwards approach.  A handful of sources were discarded as unreliable, and other individual discrepancies were resolved by inspection by DK and JM.

The main challenge in both approaches was to associate a unique set of emission lines to a unique galaxy, especially in crowded regions where a single set of {\Oiii} ($+$H$\beta$) lines may plausibly be associated to multiple galaxies.  It should be noted that limiting ourselves to objects with at least two significant line detections significantly facilitated this procedure because the line separation limits the possible range of redshift, and thus of position in the sky.  This is a significant benefit of designing the survey to search for the {\Oiii} lines, instead of, for example, H$\alpha$ with a redder filter.

\subsection{The sample of \Oiii-emitters in the J0100$+$2802 field}
\label{sec:galaxy_sample}

\begin{figure*}[t]
\centering
\includegraphics[width=6.5 in]{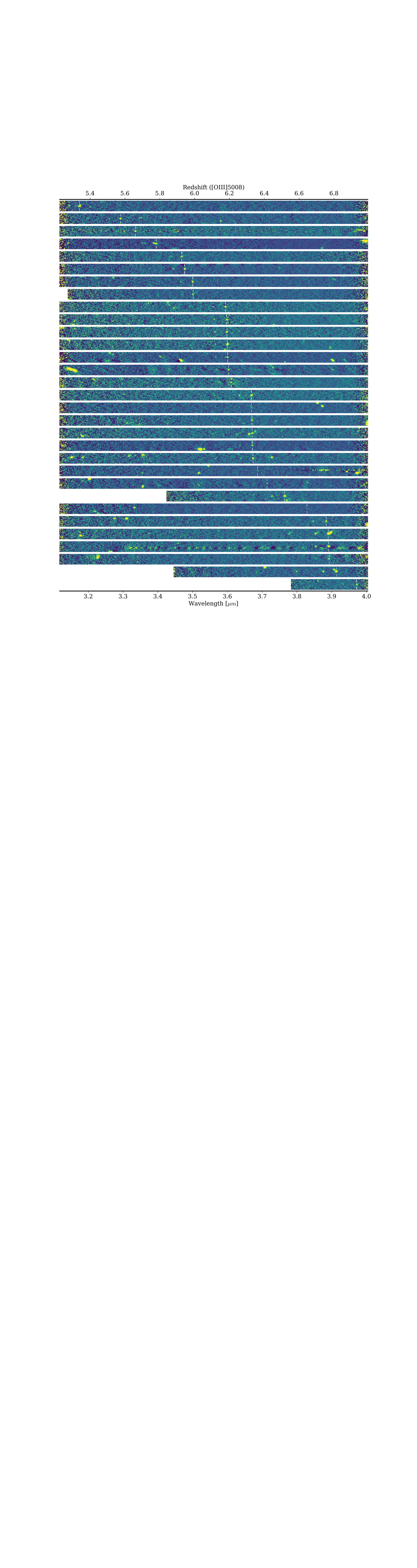}
\caption{
Examples of stacked 2D spectra of {\Oiii}-emitters, shown in order of redshift from top to bottom.  The spectra are rebinned to be flat and resampled into common wavelength grid in steps of 9.75~{\AA}~pixel$^{-1}$.  The {\Oiii}$\lambda$5008 line is marked by ticks.   Some spectra were truncated because the bluer part of the spectrum was dispersed beyond the detector coverage (see \S \ref{sec:observations}).  
\label{fig:2dspec_compilation}}
\end{figure*}

After visual inspection and reconciliation, we identified 133 \Oiii-emitting components over $z=5.33$--6.93 with at least two detected emission lines above $3\sigma$.  As detailed in Paper II, we decided to classify clumps located within $2\arcsec$ as a single ``system'', yielding a final sample of 117 unique \Oiii-emitting systems, or galaxies.   
Measurements of the emission-line fluxes were carried out using the best-quality 1D spectra that were extracted from the coadded 2D spectral images (see Paper II).  We note that they are yet subject to the $\sim5\%$-level systematic uncertainties in the absolute flux calibrations.  We also estimated the systematic redshift uncertainties to be about $\sim100~\mathrm{km~s}$, which corresponds to $\sim1$ pixel uncertainty in the wavelength calibration, from inaccuracy of the grism model.  For most cases, the systematic redshift uncertainty dominates the statistical errors.
Figure \ref{fig:2dspec_compilation} presents a representative set of stacked 2D spectra of \Oiii-emitters in order of redshift.

The full catalog of the \Oiii-emitter sample is provided in Table \ref{tb:O3_sample}.  We classified 94 sources as high-confidence sources (\texttt{CONFID}=2), and the other 23 as lower-confidence (\texttt{CONFID}=1) based on the measured S/N of the lines and any remaining ambiguity in the line--galaxy identification.  This classification of confidence was independently made by DK and JM, and any discrepancies carefully reconciled.  We estimate that these two classes are $>99\%$ and $>90\%$ reliable, respectively, for \texttt{CONFID}=2 and 1.  Further objective assessment of robustness, completeness and sample contamination will be conducted using a larger sample including future observations of the remaining EIGER fields. Our present scope only concerns early development of methods to process NIRCam data, initial data quality, and findings from the J0100$+$2802 sightline.  Throughout this paper, we use all the {\Oiii} galaxies for analysis.  Exclusion of low-confidence galaxies does not change any conclusions, except for one galaxy which is identified as associated with a metal absorption system (see \S \ref{sec:galaxy_sample_results}; the association anyway makes it therefore unlikely to be a chance spurious association with a wrong redshift).

\begin{deluxetable}{ccccc}
\tablecaption{\Oiii-emitter sample\label{tb:O3_sample}\tablenotemark{a}}
\tablehead{
    \colhead{ID}&
    \colhead{R.A.}&
    \colhead{Decl.}&
    \colhead{Redshift}&
    \colhead{CONFID\tablenotemark{b}}\\
    \colhead{}&
    \multicolumn{2}{c}{J2000}&
    \colhead{}&
    \colhead{}}
\startdata
    31 &  15.0815 &  28.0534 &  5.94 &  2 \\ 
    82 &  15.0571 &  28.0859 &  5.74 &  2 \\ 
   435 &  15.0614 &  28.0779 &  5.99 &  1 \\ 
...    &  ...     &  ...      & ... &  ... \\
 19964 &  15.0730 &  28.0115 &  6.18 &  2 \\ 
 20317 &  15.0055 &  28.0547 &  6.24 &  1 \\ 
 20379 &  15.0375 &  28.0112 &  6.77 &  1 
\enddata
\tablenotetext{a}{The coordinates and redshifts shown are tentative.  The full catalog will be made publicly available upon publication.}
\tablenotetext{b}{Reliability flag: CONFID$=2~(1)$ for higher (lower) confidence.}
\end{deluxetable}

\section{Ground-based spectroscopy of QSO J0100$+$2802}
\label{sec:qso_spectroscopy}

\begin{figure*}[t]
\centering
\includegraphics[width=7.0in]{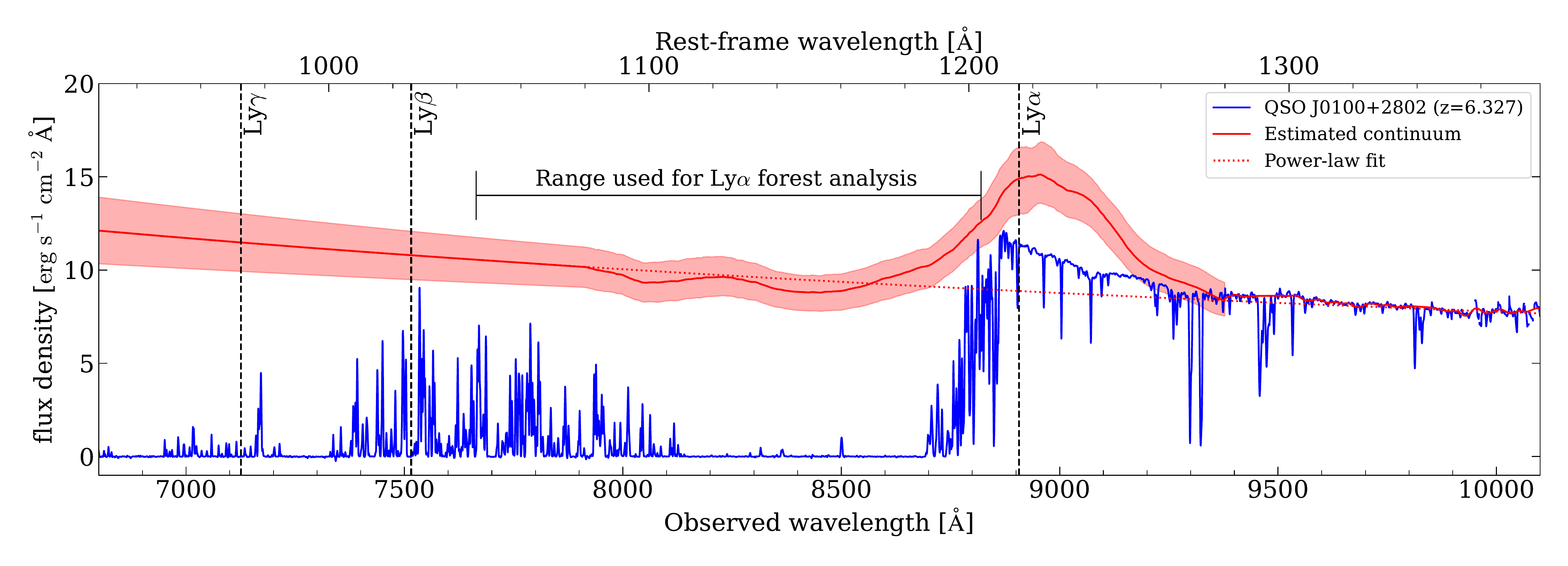}
\caption{The spectrum of QSO J0100+2802 around the Ly$\alpha$ forest region. The blue line indicates the high resolution X-Shooter spectrum and the red solid line is the estimated intrinsic continuum from a deep learning neural network with the 95\% confidence interval (shaded region). 
Blueward of rest-frame 1080~{\AA} is extended by a power-law fit to the continuum range (red dotted line). The dashed vertical
lines indicate the wavelengths of the Ly$\alpha$, Ly$\beta$, and Ly$\gamma$ lines at the quasar redshift.
\label{fig:J0100_spectrum}}
\end{figure*}

We used ground-based spectroscopic observations of QSO J0100$+$2802 to measure the transmission as a function of redshift in the Ly$\alpha$ and Ly$\beta$ forest region and to identify intervening absorption systems.  The spectroscopic data used in this paper were obtained with the X-Shooter spectrograph \citep{2011A&A...536A.105V} on the Very Large Telescope (VLT), the Folded-port InfraRed Echellette (FIRE, \citealt{2013PASP..125..270S}) on the Magellan Telescope, and the Keck/HIgh Resolution Echelle Spectrometer (HIRES; \citealt{1994SPIE.2198..362V}).

The archival X-Shooter data (ESO programme 096.A-0095) and FIRE data (PI Simcoe) were reduced, carefully corrected for telluric absorption using numerical atmospheric models, then optimally combined to maximize signal-to-noise ratios, using the \texttt{PypeIt}\footnote{\url{https://github.com/pypeit/PypeIt}} spectroscopic reduction pipeline.  A detailed description of the data reduction will be provided in a companion paper (Eilers et al.).  The HIRES data were reduced using the \texttt{makee} pipeline, and the coadded 1D spectrum was then corrected for telluric absorption using observations of a hot white dwarf spectrophotometric stardard star \citep{2019ApJ...882...77C}.

We used the X-Shooter spectrum obtained in the visible (VIS) spectrograph to measure the Ly$\alpha$ and Ly$\beta$ transmission fluxes across the redshift range of interest.  The intrinsic continuum spectrum blueward of rest-frame 1280 {\AA} was estimated from the shape of its redward spectrum using a deep neural network trained with a sample of low-redshift ($0.1 \leq z \leq 3$) quasar spectra \citep{2021MNRAS.502.3510L}.  The intrinsic spectrum that was estimated by this method extends down to the rest-frame wavelength of 1080~{\AA}, and was complemented to shorter wavelengths by a power-law fit to the continuum redward of Ly$\alpha$.  In doing so, the power-law fit was forced to continuously connect the one predicted from the deep learning network at the rest-frame 1080~{\AA}.  Figure \ref{fig:J0100_spectrum} shows the observed X-Shooter spectrum and the estimated intrinsic spectrum.  We estimated $1\sigma$ uncertainties of the predicted continuum spectrum to be $\approx 5\%$ for the Ly$\alpha$ and Ly$\beta$ forest regions that we will use for analysis.

In the redshift window covered by \Oiii ~($z=5.3\textrm{--}6.327$), we identified eight discrete intervening absorption systems using all the available spectroscopic data.  The high-resolution spectra suggest that these systems often contain multiple blended absorption components.  We used absorption-line fitting routines that generate a single hierarchical model of absorbing components and ion transitions, and project the same model into the respective spectral data space of each instrument, convolving to the appropriate spectral resolution.  
The model was optimized to fit the data from all instruments simultaneously using a Markov-Chain Monte Carlo technique implemented in the emcee package \citep{2013PASP..125..306F}.  Full details of the identification and measurements of absorption lines will be provided in dedicated companion papers (Bordoloi et al., in prep, Simcoe et al. in prep).

In this paper, the absorption systems are simply categorized into low- and high-ionization systems according to the transitions and their strengths.  Three absorption systems at absorption redshifts $z_\mathrm{abs} \approx 5.34$, 6.01 and 6.19 exhibit significant absorption by the \Civ~1548,1550 {\AA} transitions, and these are classified as high-ionization systems.  The other five systems at $z_\mathrm{abs} \approx 5.35, 5.80, 5.95, 6.11$ and 6.14 are classified as low-ionization systems and exhibit prominent neutral and/or singly-ionized metal absorption lines, and absent or very weak absorption in {\Civ} and/or {\Siiv}.  All these low-ionization systems, except the one at $z_\mathrm{abs}=5.35$, are likely to trace neutral gas as indicated by the detection of \Oi~1302~{\AA} absorption.  A particular goal of the EIGER survey is to search for host galaxies of these absorbing systems.

Here we highlight the presence of neutral gas at $z\approx 6.14$, approximately $\sim 75~\mathrm{cMpc}$ in front of the quasar, that is identified through the detection of a strong {\Oi} absorption system, previously reported in \citet{2019ApJ...882...77C} and \citet{2019ApJ...883..163B}.  
Like all luminous quasars, J0100+2802 enhances the ionization fraction of hydrogen in its local environment, as evidenced by a gradual increase in the Ly$\alpha$ transmission as the quasar's redshift is approached.  Gunn-Peterson absorption begins to saturate at $\approx 72~\mathrm{cMpc}$ in the quasar's foreground, but the presence of {\Oi} and detailed shape of the Ly$\alpha$ transmission both suggest the existence of a Damped Ly$\alpha$ absorption system at $z=6.14$, which would fully absorb Lyman continuum radiation from the quasar along the line of sight. 
This implies that ionizing radiation from the quasar will be fully shielded from the foreground IGM below this redshift.  This motivates us to use this redshift as a natural boundary to divide the regions with and without the effect of the quasar's ionizing radiation.  In this work, we define the region just above this division redshift (up to the quasar's redshift) as the so-called ``near zone'' of the quasar\citep[e.g.,][]{2010ApJ...714..834C,2015ApJ...801L..11V,2017ApJ...840...24E}.

\section{Galaxies along the QSO J0100$+$2802 sightline}
\label{sec:galaxy_sample_results}

\begin{figure*}[t]
\centering
\includegraphics[width=7.0 in]{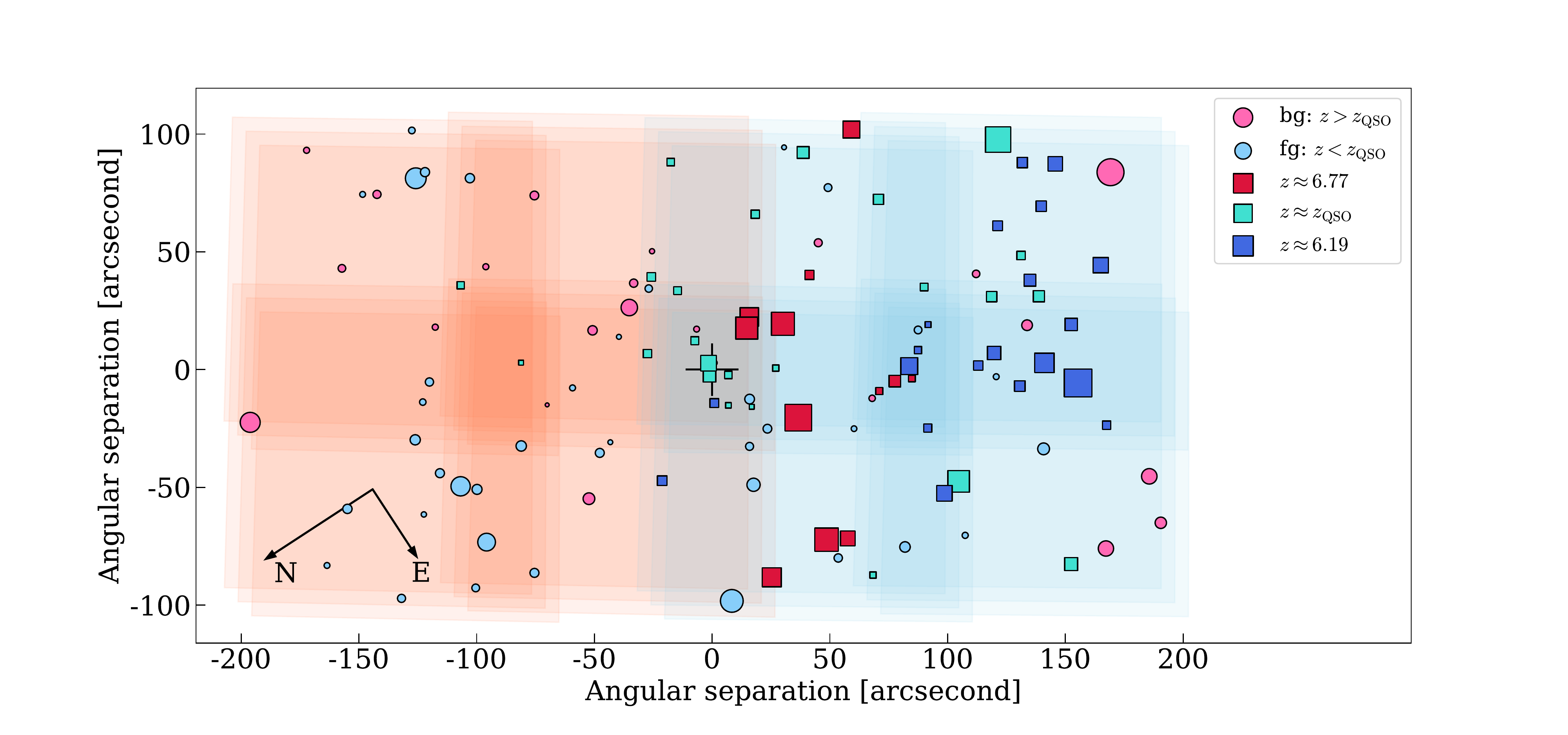}
\caption{
The sky distribution of the detected \Oiii-emitting galaxies, with sizes coded according to the relative {\Oiii} luminosity.  Galaxies identified as being within the three strongest overdensities are marked in squares.  Other galaxies are shown by circles and color-coded according to whether they are foreground (light blue) or background (pink) of the quasar.  The detector fields-of-view of the individual exposures are shown so that the color intensity corresponds to the effective exposure time, with the red and blue colors for the Module A and B detectors, respectively.  The axes values are the angular separations from the position of QSO J0100$+$2802, which is represented by the large cross symbol.
\label{fig:O3_map}}
\end{figure*}

\begin{figure*}[t]
\centering
\includegraphics[width=7.0 in]{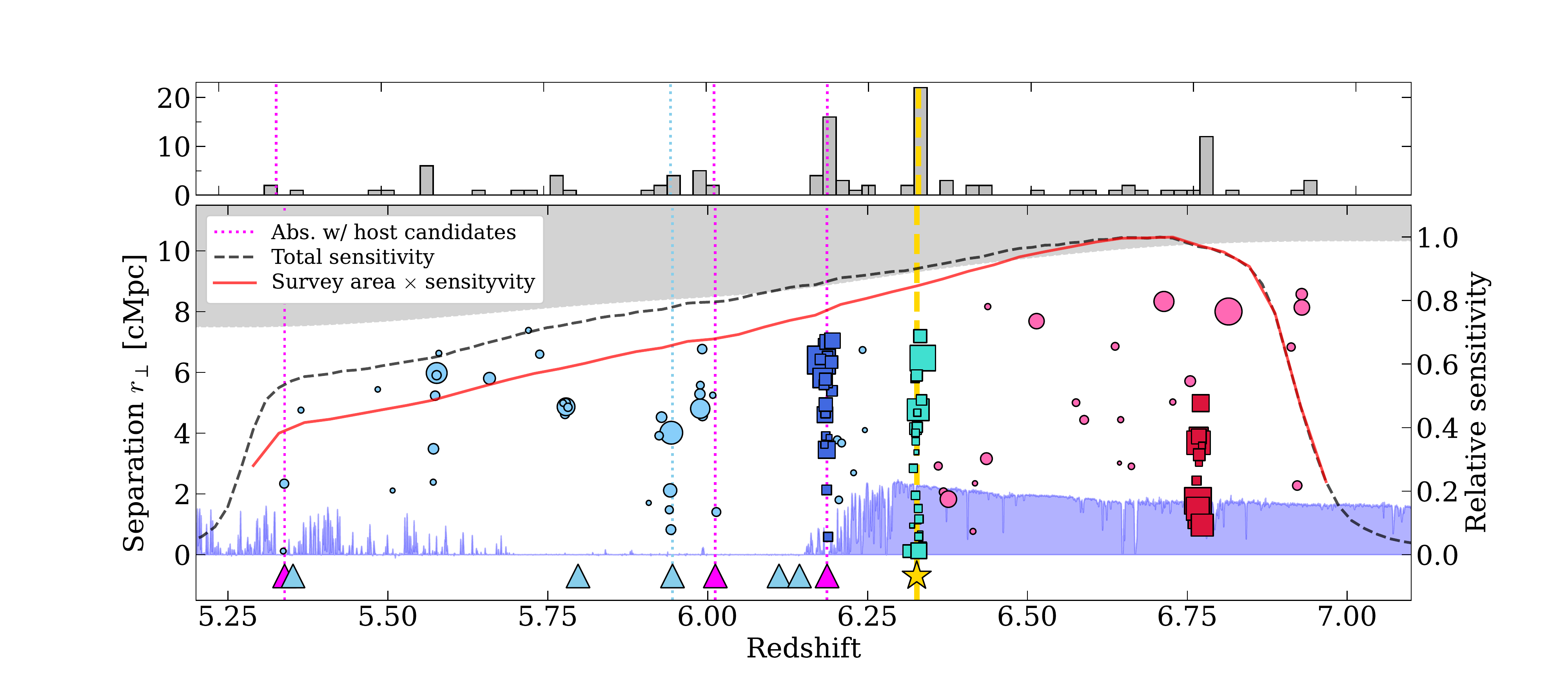}
\caption{
The transverse separation of the \Oiii-emitters from the quasar sightline are shown as a function of redshift (main panel) with the redshift histogram (subpanel).
The symbols follow the same color- and size-coding as in Figure \ref{fig:O3_map}.
For reference, the observed QSO J0100$+$2802 spectrum is shown in arbitrary flux units, where the wavelength is translated into the redshift for Ly$\alpha$.
The gold star and vertical dotted line mark the quasar redshift ($z_\mathrm{QSO}=6.327$). The triangles indicate metal absorption systems in the two classes of low-ionization (cyan) and high-ionization systems (magenta).  The vertical dotted lines mark the absorption systems for which host galaxy candidates were identified.  Essentially, all galaxies within $200~pkpc$ of the sightline have an associated metal absorption system.  The gray shaded region indicates a forbidden area, i.e., where an object is too far from the sightline to be within the field of view covered by the area of the detector.  The black dashed curve gives an idea of the redshift sensitivity for the {\Oiii} line because of the filter transmission curve and other effects.  The red solid curve then multiplies this by the fraction of the field of view over which an {\Oiii} line at that redshift will fall onto the detector.  The scale of these curves is arbitrary and normalized to their peaks.
\label{fig:O3_redshift_dist}}
\end{figure*}

We here provide an overview of this first sample of 117 {\Oiii}-emitting galaxies and some of the most immediate findings.  These galaxies have \Oiii$\lambda$5008 luminosities $ \log L_\mathrm{[OIII]5008}/(\mathrm{erg~s^{-1}}) \approx {42}$--$43.5$, UV magnitudes $M_\mathrm{UV} \approx -18$ to $-22$, stellar mass $\log_{10} M_\mathrm{star}/M_\odot \approx 7$--9.5 (median $10^{7.8}~M_\odot$), and the average star formation rates of $\approx 11~M_\odot~\mathrm{{yr}^{-1}}$ (estimated from the combined spectrum of the entire sample).  The measurements of  physical quantities and the detailed characterization of the {\Oiii}-emitters are presented in Paper II.

As shown in Figure \ref{fig:O3_map}, we detected more galaxies (59\%) in the right-hand side region mostly covered by Module B than the other side covered by Module A.  Given the less sensitivity in Module B, this should reflect the real spatial distribution of the galaxies, rather than being an artificial bias.

\subsection{Overdensities along the line of sight}

Figures \ref{fig:O3_map} and \ref{fig:O3_redshift_dist} show, respectively, the spatial distribution and the transverse separations of the \Oiii-emitters from the quasar sightline as a function of redshift along the line of sight.  The galaxies show a strongly clustered redshift distribution with sharp overdensities at particular redshifts.  
In particular, three prominent overdensities are clearly visible at $z\approx6.19$, 6.33, and 6.78, respectively, in front of, precisely at, and behind the quasar redshift.  Galaxies within these three specific structures are highlighted in the figures.  These member galaxies were defined by grouping galaxies in redshift space with a linking length of $200~\mathrm{km~s^{-1}}$, which is comparable to the velocity dispersion of the two lower-redshift overdensities.
 
\begin{figure*}[t]
\centering
\includegraphics[width=7.0in]{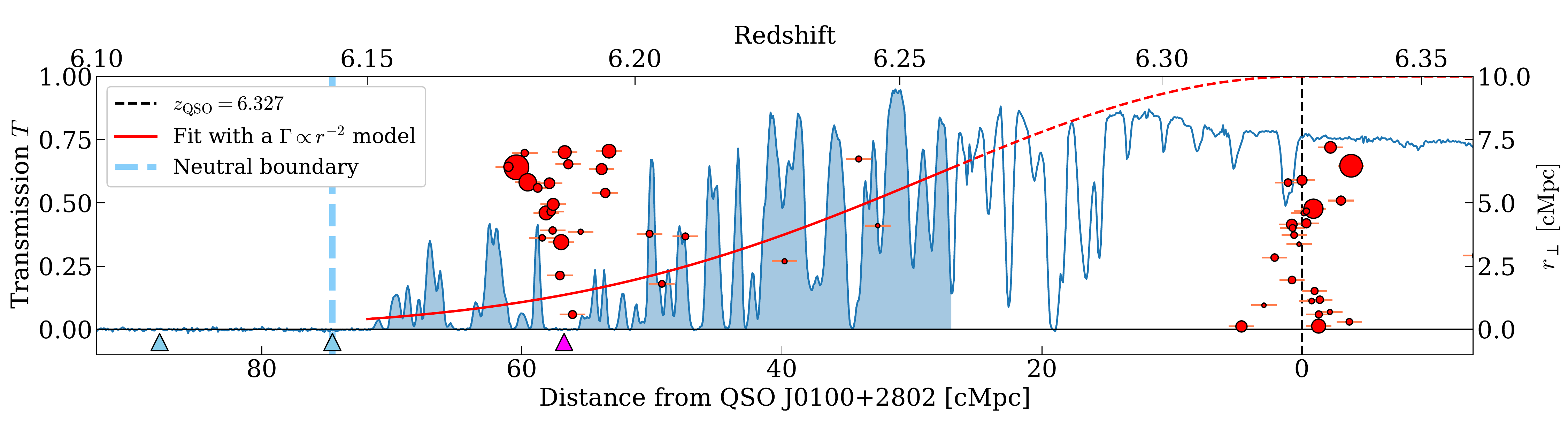}
\caption{The Ly$\alpha$ transmission in the quasar near zone as a function of radial comoving distance from {\QSO}.   The vertical dashed line indicates the location of the quasar ($z_\mathrm{QSO}=6.327$).
The red circles show the \Oiii-emitters, size-coded according to the relative {\Oiii} luminosity.  The horizontal error bars correspond to the nominal redshift uncertainties of $\pm100~\mathrm{km~s^{-1}}$.  Strong overdensities are evidently seen at $z\approx6.19$, within the near zone, and at $z\approx6.33$, coincident with the quasar redshift.  
The red curve indicates a characterization of the near-zone Ly$\alpha$ transmission profile, assuming that the dominant radiation field $\Gamma \propto r^{-2}$, fitted to the transmission data across the shaded region (see Section \ref{sec:Lyaforest}).  The triangles mark the metal absorption systems in the same way as in Figure \ref{fig:O3_redshift_dist}.  The vertical blue dashed line represents a neutral system which, it can be safely assumed, blocks all ionizing radiation from the quasar from penetrating to lower redshifts.
\label{fig:qso_nearzone}}
\end{figure*}

The $z \approx 6.19$ overdensity contains 20 galaxies within $6.176\le z\le6.195$ with the velocity dispersion $\sigma_v = 212~\mathrm{km~s^{-1}}$.  The galaxy spatial distribution is biased towards the right-hand side of the field of view (Figure \ref{fig:O3_map}). Figure \ref{fig:qso_nearzone} compares the distribution of {\Oiii} sources and the Ly$\alpha$ transmission fluxes.
This structure lies $\approx 57~\mathrm{cMpc}$ in front of {\QSO}, i.e., within the near zone with strong transmission that is defined to be down to the ``natural boundary'' of neutral gas traced by {\Oi} absorption system at $z=6.14$ (see \S \ref{sec:qso_spectroscopy}).  
Moreover, this overdensity, or at least the member galaxies of it, appear to physically connect to the high-ionization metal-absorption system at $z=6.19$.  The local effect of these galaxies on the surrounding IGM is examined in Section \ref{sec:Lyaforest}.

The second overdensity, containing 24 galaxies with $\sigma_v=197~\mathrm{km~s^{-1}}$, is exactly at the redshift of \QSO, as shown in Figure \ref{fig:qso_nearzone}.  
This is indeed the largest overdensity, in terms of the number of member galaxies, along the line of sight, and implies that the SMBH that powers this most luminous quasar formed in a high overdensity of galaxies.  The centroid of the galaxy positions is about $\approx 40\arcsec$, or $\approx1.6~\mathrm{cMpc}$, away from the quasar position.  On the other hand, Figure \ref{fig:qso_nearzone} shows some evidence that the velocity dispersion becomes larger at very small separations from the quasar.  The velocity dispersion within $r<1~\mathrm{cMpc}$ of the quasar is $\sigma_v=300~\mathrm{km~s^{-1}}$, twice as higher as the $\sigma_v=147~\mathrm{km~s^{-1}}$ for $r>1~\mathrm{cMpc}$ ($\sigma_v=94~\mathrm{km~s^{-1}}$ if limited to $1<r/\mathrm{cMpc}<5$).  This may reflect a Finger-of-God effect from virialization of the system.  The nature of this structure will be examined in companion papers (Eilers et al., in prep, Mackenzie et al. in prep).

The third overdensity, with 12 member galaxies and $\sigma_v=100~\mathrm{km~s^{-1}}$, is behind the quasar, and comparable in redshift to the most distant proto-clusters previously reported \citep{2021NatAs...5..485H,2022MNRAS.511.6042E}.  
Intriguingly for their very high redshifts, the median \Oiii$\lambda$5008 luminosity ($\sim2.8\times10^{43}~\mathrm{erg~s^{-1}}$) of the member galaxies in this structure is $>6$ times higher than the average of the other galaxies in the sample ($\sim4.2\times10^{43}~\mathrm{erg~s^{-1}}$).

\subsection{Associations with metal absorption systems}

The deep NIRCam WFSS observations enable a high-completeness search for host galaxies of the known metal absorption systems along this line of sight.  Using the {\Oiii}-emitter sample, we identified host galaxy candidates within $\pm 500~\mathrm{km~s^{-1}}$ and 300 physical kpc (pkpc) for four absorption systems ($z_\mathrm{abs}=5.34, 5.95, 6.01, 6.19$) out of the eight known systems.  
In Figure \ref{fig:O3_redshift_dist}, the absorption systems with host candidates are marked by vertical dotted lines.  These candidate associations in fact represent {\it all} the sample galaxies that lie within 200~pkpc of the sightline within the redshift range studied: in other words, if a galaxy is within 200~pkpc of the quasar sightline then it has a metal absorption line associated with it.

As noted above, the galaxy associated with the metal absorber at $z=5.33$ was a priori classified as being in the second confidence class (\texttt{CONFID}=1) mostly because this object is at the bluer edge of the sensitivity curve which makes it hard to assess the {\Oiii}~4960-to-5008 ratio.  However, this coincidence, with the remarkably small impact parameter of (only) 19~pkpc and a velocity offset $\approx 120~\mathrm{km~s^{-1}}$, further supports that this is truly a high-$z$ \Oiii-source and that it is associated with the absorber.

Interestingly, absorber-host candidates were identified for all three of the high-ionization systems, including the $z_\mathrm{abs}=6.19$ system coincident with the aforementioned prominent overdensity, but only for one out of the five low-ionization systems.  This could reflect the difference in the galaxy populations and/or physical conditions in the absorption systems at different ionization conditions.  A more detailed study of these absorber--galaxy associations, including those identified at lower redshifts via other emission lines such as H$\alpha$ will be presented in companion papers (Bordoloi et al., in prep; Simcoe et al., in prep).

In the remainder of this paper, we focus on the subsample of 59 \Oiii-emitters which are located within the redshift range where the transmission of Ly$\alpha$ and Ly$\beta$ along the line of sight may be probed by the quasar spectrum.

\section{Correlation of galaxies with IGM transmission}
\label{sec:Lyaforest}

\begin{figure*}[t]
\centering
\includegraphics[trim={1cm 0 0 0}, width=7.2in]{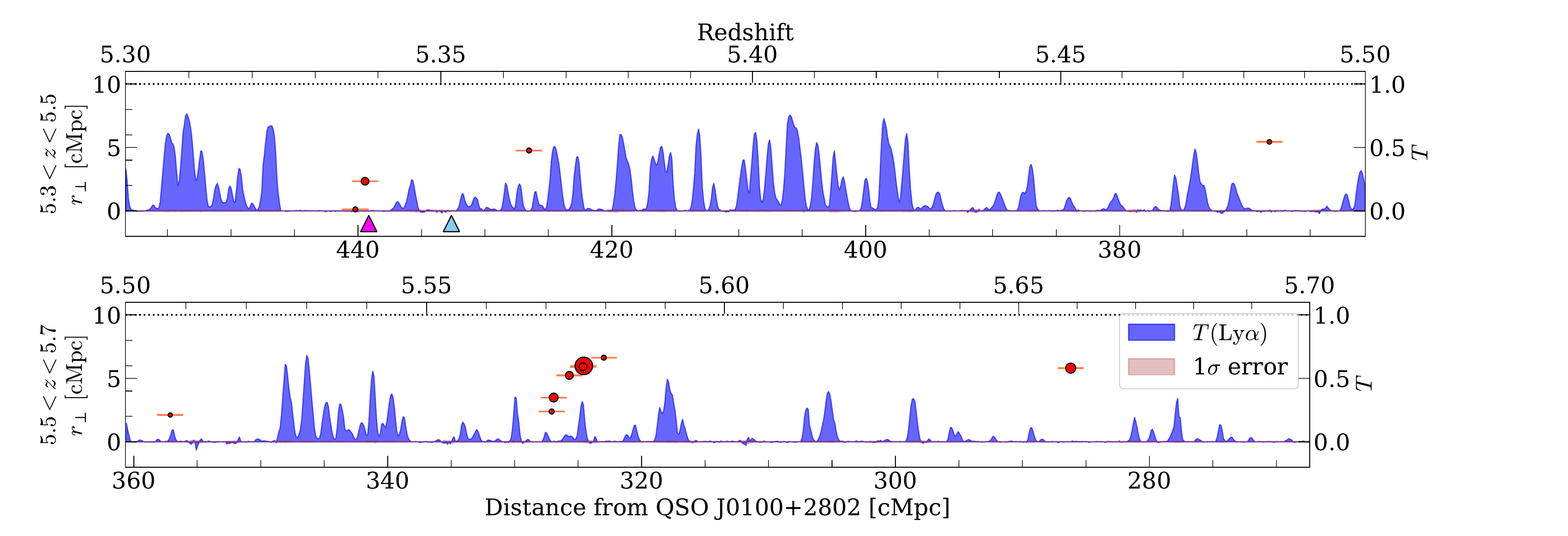}
\caption{
Ly$\alpha$ transmission $T$ along the line of sight to QSO J0100$+$2802 and the comoving transverse separation $r_\perp$ of the galaxies from the quasar sightline in the redshift range $5.3 < z < 5.7$, the lowest in our study.  The red circles show the radial distances to the detected \Oiii-emitters from the sightline with different sizes proportional to the {\Oiii} luminosity.  The horizontal error bars correspond to the nominal redshift uncertainties of $\pm100~\mathrm{km~s^{-1}}$.  The triangles at the bottom of each panel mark the positions of  high- (magenta) and low-ionization (cyan) metal absorption systems.  In all the panels, the $x$ and $y$-axes are scaled equally in comoving distance.
\label{fig:Lyaforest_O3_1}}
\end{figure*}

\begin{figure*}[t]
\centering
\includegraphics[trim={1cm 0 0 0}, width=7.2in]{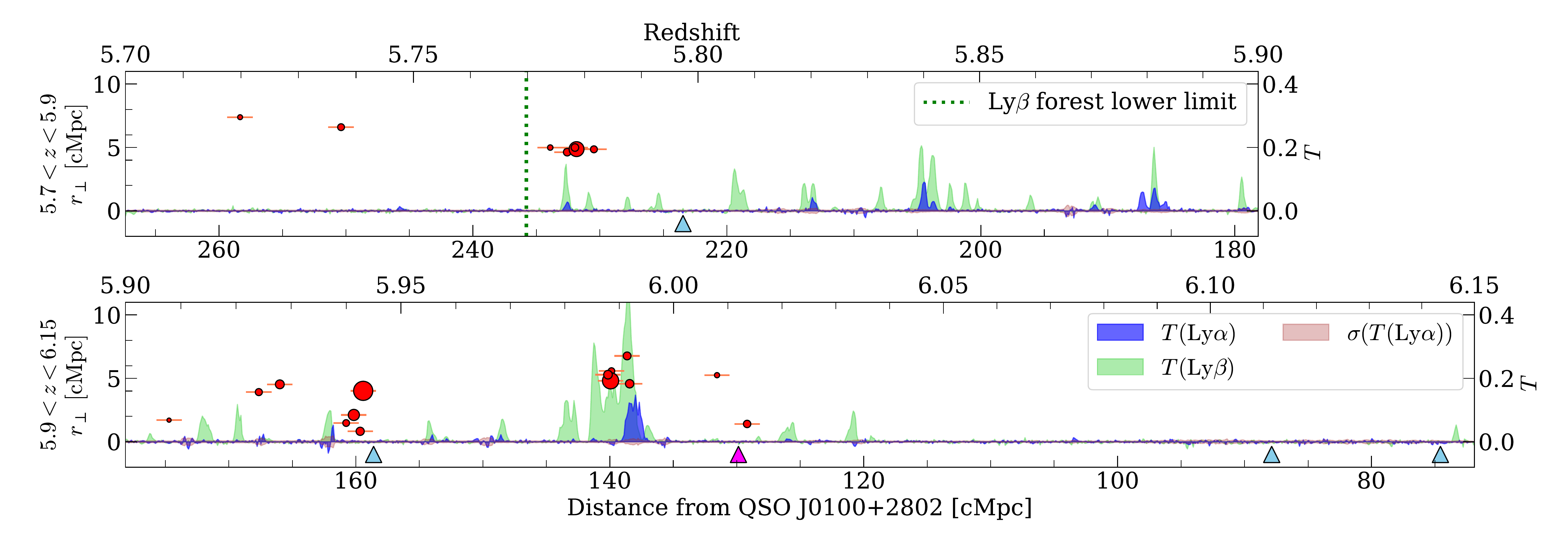}
\caption{
Same as Figure \ref{fig:Lyaforest_O3_1}, but for the higher redshift range of $5.7 < z < 6.15$.  The corresponding Ly$\beta$ transmission, which is also available in this redshift interval, is also shown in green.  The vertical dotted line indicates the lower limit of the Ly$\beta$ forest range used for analysis ($z_\mathrm{abs} = (1+6.14) \lambda_\gamma/\lambda_\beta-1 = 5.77$; see text).
\label{fig:Lyaforest_O3_2}}
\end{figure*}

In this section, we investigate the  relation between galaxies and the surrounding IGM during the later stages of reionization by measuring the cross-correlation between the location of \Oiii-emitting galaxies and the transmission of the IGM in Ly$\alpha$ and Ly$\beta$.

As shown in Figures \ref{fig:J0100_spectrum} and \ref{fig:O3_redshift_dist}, the Ly$\alpha$ transmission varies strongly across the range of redshifts accessible.  However, we may identify the three different regimes: 1) $5.3 < z < 5.7$, where the Ly$\alpha$ transmission is quite strong and much more ubiquitous than at higher redshifts; 2) $5.7 < z < 6.14$, a relatively dark Gunn-Peterson trough in which Ly$\alpha$ transmission is only seen in isolated discrete transmission spikes; 3) $z \gtrsim 6.15$, where the transmission increases towards the quasar, due to the ionizing radiation from the quasar itself.  The division between the first two regimes is somewhat arbitrary.  The second division is motivated by the ``natural boundary'' produced by the existence of the neutral medium traced by the $z=6.14$ {\Oi} metal absorption system, as discussed in \S \ref{sec:galaxy_sample_results}.  This system should block ionizing radiation from the quasar from penetrating to all lower redshifts along the line of sight.

Figures \ref{fig:Lyaforest_O3_1} and \ref{fig:Lyaforest_O3_2} show a zoom-in along the line of sight in the first two redshift regimes identified above.
For each galaxy, the comoving transverse separation is shown vertically, as a function of the comoving distance from the quasar with the same scaling, as computed (assuming the Hubble flow) from the observed redshifts indicated along the upper horizontal axis.  The $y$-axis on the right-hand side shows the fractional transmitted Ly$\alpha$ flux where the observed wavelength is converted into absorption redshift.  Note that the transmitted fluxes are subject to 5\%-level systematic uncertainties in predicting the intrinsic continuum spectrum.  Figure \ref{fig:qso_nearzone} showed the equivalent for the third ``near zone'' redshift regime.

Comparing the Ly$\alpha$ transmission to the distribution of the identified galaxies, we can immediately see clear coincidences between galaxies and transmission spikes at $z \approx 5.785$ and 5.99.  The transmission is however likely to be suppressed nearer to the galaxies, as seen e.g., at $z=5.335$ (Figure \ref{fig:Lyaforest_O3_1}), 5.94 (Figure \ref{fig:Lyaforest_O3_2}), and around the $z=6.19$ overdensity within the quasar near zone. 
It should also be noted that Ly$\alpha$ transmission is not seen near the known metal-absorber systems (and thus near their host candidate galaxies).  All these findings accumulate evidence that star-forming galaxies directly impact on the surrounding IGM, but in a complex manner.

\subsection{Average transmission around galaxies}

\begin{figure}[t]
\centering
\includegraphics[trim={1cm 0 0 0}, width=3.4in]{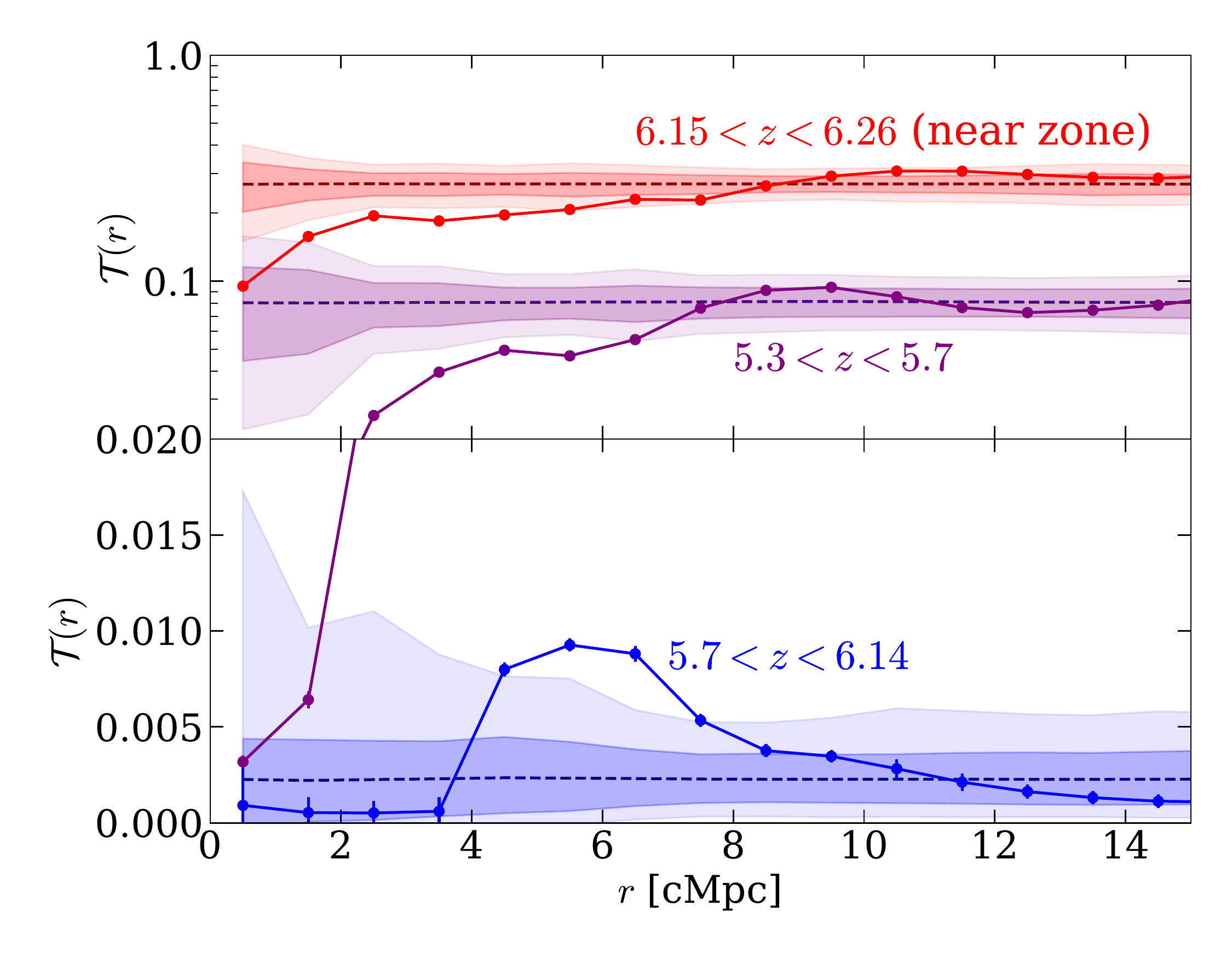}
\caption{
The average Ly$\alpha$ transmission as a function of distance from the \Oiii-emitter galaxies in the three redshift regimes $5.3<z<5.7$, $5.7<z<6.14$, and $6.15<z<6.24$.   The upper and lower panels show the values with log and linear scales to accomodate the wide range in the data.   The dashed lines and the shaded regions indicate the mean and the central 68 and 95 percentiles of the probability distribution obtained when the transmission along the line of sight is randomly scrambled (see text for details).   At the lowest redshifts, the transmission increases monotonically to larger distances.  At slightly higher redshifts, as the EoR is approached, the transmission curve shows a pronounced peak around 5--6~cMpc.  Finally, in the quasar near zone, where the ionizing radiation from the quasar dominates, the transmission reverts to the earlier monotonic behaviour.
\label{fig:Tavg}}
\end{figure}


To quantify the effects of galaxies on the IGM, we measured the average Ly$\alpha$ transmission, denoted  $\mathcal{T}(r)$, as a function of comoving distance $r$ from the \Oiii-emitters in these three redshift regimes ($5.3<z<5.7$, $5.7<z<6.14$, and $6.15<z<6.26$).

Figure \ref{fig:Tavg} shows the galaxy--Ly$\alpha$ transmission cross-correlation, or the transmission curves $\mathcal{T}(r)$.  
In each of the two lower redshift regimes, we took out the effect of (relative) redshift evolution within each redshift range by dividing by the cosmic mean transmission as a function of redshift \citep{2018ApJ...864...53E} and then multiplying back by the average within the redshift range in question (to revert back to the absolute scale):
\begin{equation}
\tilde{T} (z_\mathrm{Ly\alpha}) = \frac{T(z_\mathrm{Ly\alpha})}{T_\mathrm{mean} (z_\mathrm{Ly\alpha})}  \times \frac{\int_{z_\mathrm{min}}^{z_\mathrm{max}} T_\mathrm{mean} (z) dz}{z_\mathrm{max}-z_\mathrm{min}},
\end{equation}
where $T_\mathrm{mean}$ is the cosmic mean transmission and $\tilde{T}$ is the product of this step that is used for calculating the transmission curve $\mathcal{T}(r)$.

In the near-zone regime where the ionizing radiation is dominated by the quasar, we adopted a modeled average transmission profile as a function of distance from the quasar, instead of the cosmic mean.  This is shown in Figure \ref{fig:qso_nearzone}.
The model profile assumes that the surrounding IGM density is constant and that the radiation field $\Gamma$ scales as $r^{-2}$ where $r$ is the distance from the quasar.  This produces $T \propto \exp(- a r^2)$ when the neutral fraction is substantially small.  Here the (constant) parameter $a$, to be determined from the data, reflects the combination of the IGM density and the luminosity of the ionizing radiation.

Because we want to see if the IGM transmission is affected by nearby galaxies, we compared the observed transmission curves, $\mathcal{T}(r)$, with Monte-Carlo realizations using random transmission spectra.  The random data were generated by partitioning the real quasar spectrum into 0.7~cMpc-width bins and shuffling their order within the redshift range.  This reshuffling bin-size is comparable to the instrumental resolution and thus also comparable to the narrowest possible width of discrete transmission spikes.  We applied Gaussian smoothing with $\sigma=2~\mathrm{cMpc}$ to both real and random transmission spectra before measuring the average transmission around the galaxies.  We limited our analysis within $r<15~\mathrm{cMpc}$, i.e., $\sim2\times$ the maximum transverse distance covered by our data ($\sim7\mathrm{cMpc}$ at $z\sim6$) in the survey area.

Figure \ref{fig:Tavg} illustrates that the overall level of Ly$\alpha$ transmission, observed at larger $r$ in the random realization, varies strongly between the redshift ranges.  This reflects the global evolution of the cosmic ionizing radiation field with redshift, plus the strong near-zone ionization effect due to the quasar in the highest redshift regime.

At the lowest redshift range, the transmission $\mathcal{T}(r)$ is very low near to galaxies, and monotonically increases with distance away from them, reaching an asymptotic value at $r\sim8~\mathrm{cMpc}$.  This monotonic increase in transmission away from galaxies is similar to what is observed at lower redshifts \citep[e.g.,][]{2014MNRAS.442.2094T,2017MNRAS.471.2174B,2020ApJ...903...24M,2020MNRAS.499.1721C,2021ApJ...909..117M}.
This monotonic increase in $\mathcal{T}(r)$ can be interpreted to be reflecting the ionization of the IGM by a more or less uniform cosmic ionizing background, coupled with a higher gas density surrounding the galaxies.

In contrast, in the intermediate redshift regime, the average transmission $\mathcal{T}(r)$ first appears to increase at small distances, likely due to the density effect,  but then $\mathcal{T}(r)$ clearly peaks at $\sim 5\textrm{--}6~\mathrm{cMpc}$ away from galaxies before strongly declining towards the ``average'' value at still larger distances.  This dramatic change in behavior is evidence that the local ionizing radiation from nearby galaxies is dominating over the cosmic ionizing background on these 5~cMpc scales, because of the reduction in the latter, and that the local radiation from these galaxies, or fainter galaxies that are similarly distributed, is responsible for increasing the ionization of the surrounding IGM and thus producing the observed transmission spikes. Establishing whether the ionization is really due to the galaxies that we actually detect, or to undetected fainter sources similarly distributed, may require amongst other things an estimate of the ionizing photon escape fraction of the detected sources.

Finally, in the highest-redshift range, in the quasar near-zone, the behaviour of $\mathcal{T}(r)$ appears to qualitatively revert to the monotonic trend with distance.  The average transmission increases with $r$ out to$\sim 8~\mathrm{cMpc}$).  While it should be noted that the cross-correlation signal is dominated by the single strong overdensity at $z=6.19$, this behaviour is similar to what is seen in the lowest redshift range (without the quasar), and can be consistently explained as a natural consequence of denser gas around galaxies in a region dominated by a large-scale ionizing radiation field, in this case coming from the exceptionally luminous quasar.  This situation may be more similar to that of lower redshifts, $z\sim 4.5$, where the cosmic ionizing background, and thus transmission, are higher.

As stressed earlier, the EIGER galaxy sample is complete in the sense that all galaxies that were detectable in the survey area, above our flux limit (but with some spatial variations), should indeed appear in our galaxy catalogue.  It is clear that some transmission spikes appear along the line of sight without any associated galaxies.  
Unless there are galaxies similar to those detected in our survey lying just outside the field of view (whose minimum dimension is only $\sim 3.8~\mathrm{cMpc}$ from the sightline), the absence of detected galaxies associated with these spikes suggests that they are caused by fainter galaxies.

\subsection{Ly$\beta$ transmission}

\begin{figure}[t]
\centering
\includegraphics[trim={8mm 0 0 0}, width=3.6in]{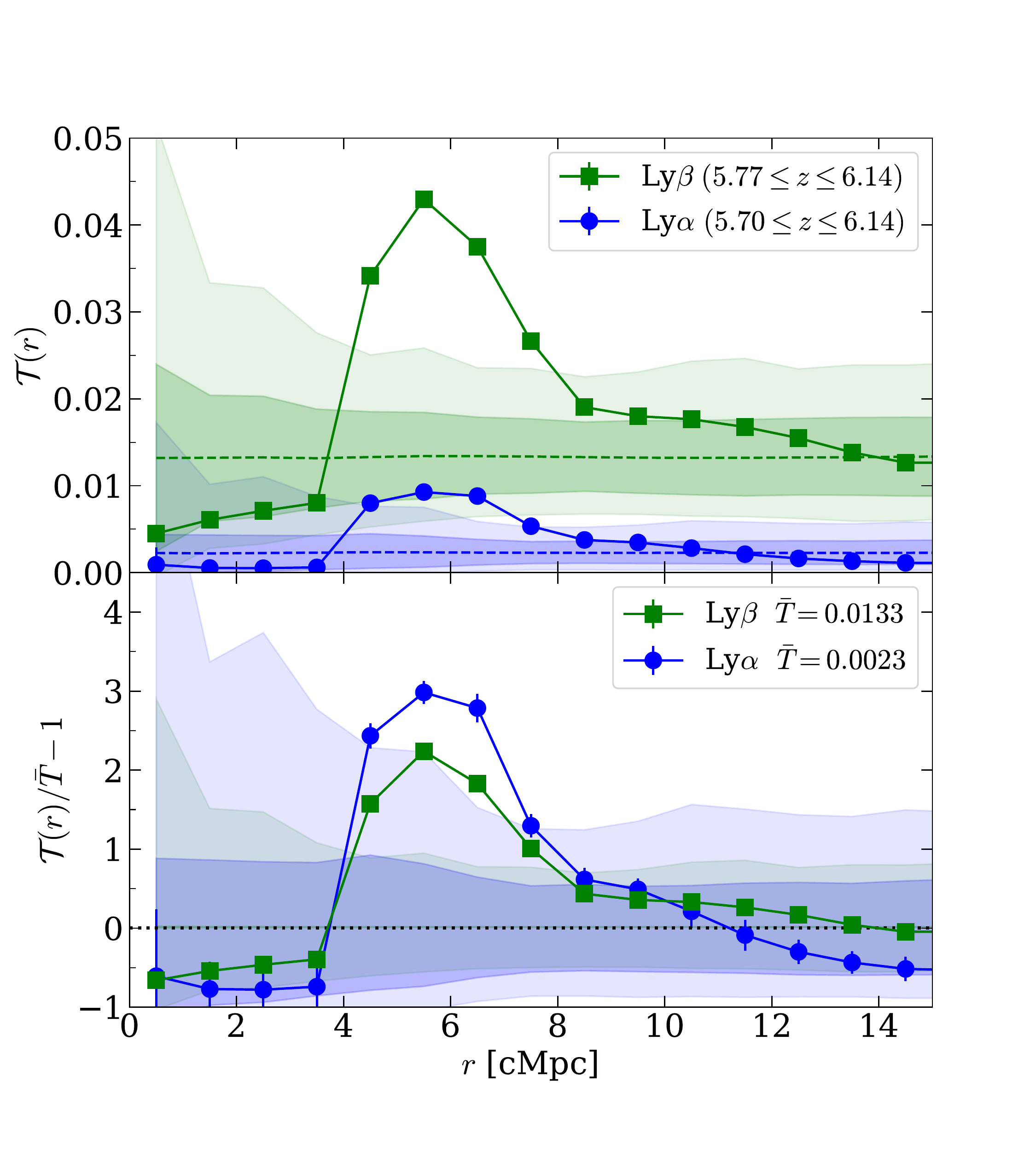}
\caption{
Upper panel: the average transmission of Ly$\beta$ (green squares) as a function of distance from the galaxies, at $5.77 < z < 6.14$, compared to that of Ly$\alpha$ in the same redshift range (blue circles, taken from Figure \ref{fig:Tavg}).
In practice, the Ly$\beta$ transmission represents a lower limit since the spectrum is affected by lower-redshift Ly$\alpha$ and higher-redshift Ly$\gamma$ absorption, but the effect should be independent of distance.
Strong enhancement of transmission at 5--6~cMpc is seen both in Ly$\alpha$ and Ly$\beta$.
Lower panel: The transmission curves normalised by the mean transmission $\bar{T}$ at that redshift, which should statistically remove the effect of contaminating Ly$\alpha$ and Ly$\gamma$ absorption on the Ly$\beta$ transmission.  This comparison emphasizes the similarity of the normalised transmission curves in Ly$\alpha$ and Ly$\beta$.  
\label{fig:Tavg_Lyb}}
\end{figure}

To add confidence in this analysis, we also examined the transmission of Ly$\beta$ in the intermediate redshift range (where the Ly$\beta$ forest is still observationally accessible).  Note that we will only see transmission in the Ly$\beta$ forest if there is transmission in both Ly$\beta$ at the redshift in question and in Ly$\alpha$ at some lower redshift.  Thus (absorption by) Ly$\alpha$ may suppress the appearance of Ly$\beta$ transmission but cannot enhance it.
Figure \ref{fig:Lyaforest_O3_2} shows that the inferred Ly$\beta$ transmission is clearly coincident (i.e. occurs at the same redshifts) with the Ly$\alpha$ transmission but also emerges where the Ly$\alpha$ absorption has saturated, because of its lower oscillator strength.

As noted above, the observed transmission flux gives lower limits on the Ly$\beta$ transmission because it is affected by the lower-redshift Ly$\alpha$ absorption.  There will be additional contamination from higher-redshift Ly$\gamma$ absorption at $z_\mathrm{abs} < 5.95 \approx (1+z_\mathrm{QSO})\lambda_\gamma/\lambda_\beta-1$ where $\lambda_\gamma=972.54$~{\AA} and $\lambda_\beta = 1025.72$~{\AA}.
We therefore adopted the redshift range down to $z_\mathrm{abs} = 5.77 \approx (1+6.14)\lambda_\gamma/\lambda_\beta-1$ where the Ly$\gamma$ absorption should not be very strong as this lies within the quasar near zone.  Indeed, we do see Ly$\beta$ transmission down to this redshift.

Figure \ref{fig:Tavg_Lyb} shows the average inferred (i.e. lower limit) Ly$\beta$ transmission curve $\mathcal{T}(r)$ around galaxies, compared with that derived above for Ly$\alpha$ in the same redshift regime.  
Similarly to the analysis of Ly$\alpha$ transmission, we removed the effect of the cosmic evolution within the redshift interval in question, using the cosmic mean Ly$\beta$ transmissivity and the mean foreground Ly$\alpha$ absorption at $z^\prime_\mathrm{abs} = (1+z_\mathrm{abs})\lambda_\beta/\lambda_\alpha-1$ where $\lambda_\alpha=1215.67$\AA \citep{2006AJ....132..117F,2018ApJ...864...53E}.  Furthermore, the effects of Ly$\alpha$ and Ly$\gamma$ absorption at different unrelated redshifts (thus unrelated distances) are taken out by normalizing the transmission by the mean transmission within the redshift range in question, as shown in the lower panel.  The normalized transmission curves show quite similar behaviors in Ly$\alpha$ and Ly$\beta$, and the coincident, strong enhancements in transmission at $r\sim5~\mathrm{cMpc}$ is seen in both in Ly$\alpha$ and Ly$\beta$.  
This strengthens the evidence for local ionizing effects, operating at distances of around 5~cMpc, as being responsible for producing the observed transmission spikes.

\subsection{Comparison with the \textsc{THESAN} simulation}

\begin{figure}[t]
\centering
\includegraphics[trim={8mm 0 0 0}, width=3.7in]{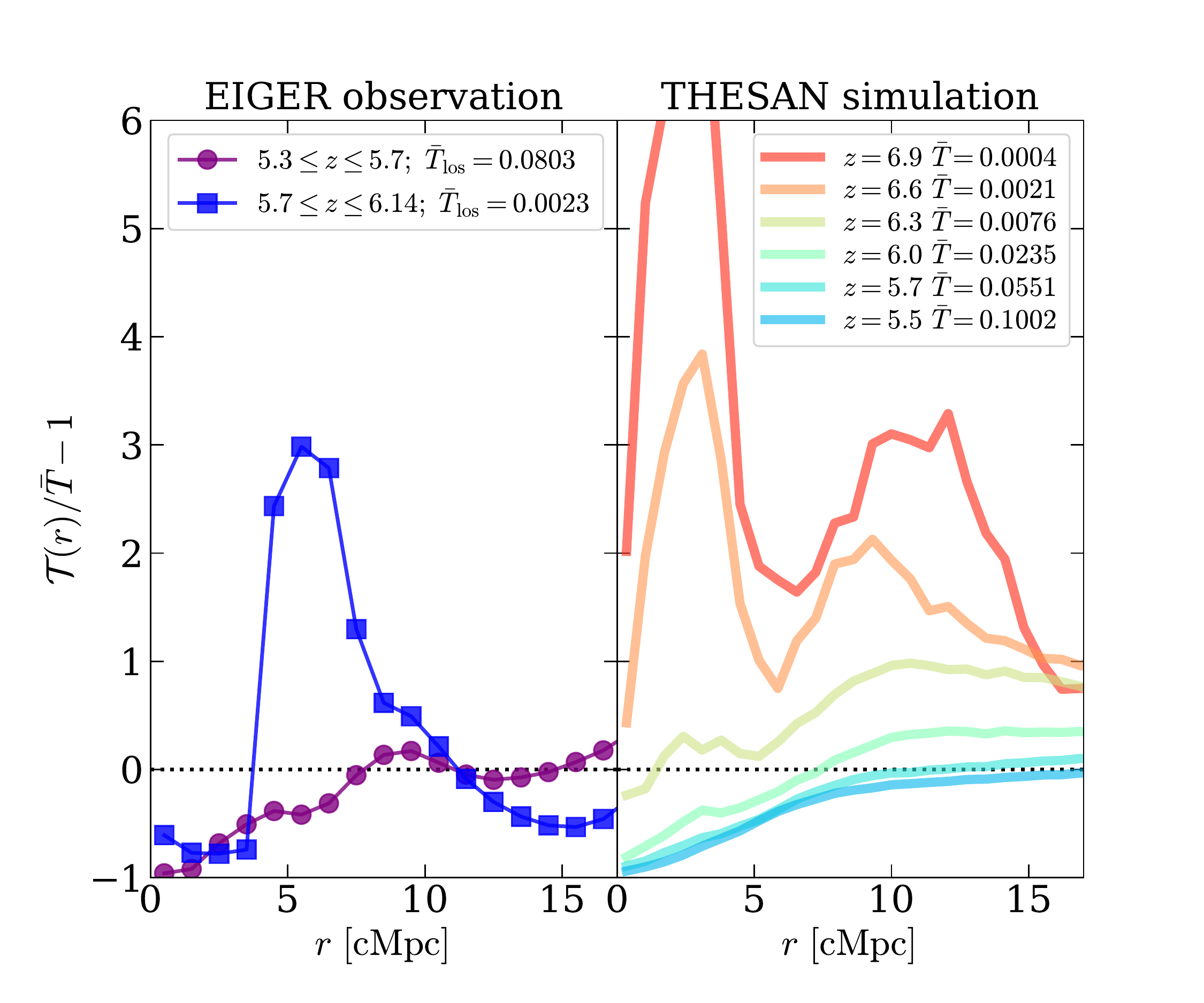}
\caption{
Comparison of the normalized transmitted fluxes that we have observed outside of the quasar near-zone (left panel) with the predictions from the hydrodynamic \textsc{THESAN} simulation (right panel). The average transmission in each case is shown in the legend.
The monotonic trend with distance from the galaxies that is observed at $z=5.3\textrm{--}5.7$ is similar to that in the simulation at $z=5.5$--5.7.  
The strong enhancement of transmission at distances of around 6 cMpc from galaxies at $z=5.9\textrm{--}6.14$ is qualitatively similar to that seen in the simulation at $z \gtrsim 6.6$, when the mean transmissivity $\bar{T}$ is actually comparable, but is at somewhat different distances  frm the galaxies ($r \le 3~\mathrm{cMpc}$). 
\label{fig:comp_THESAN}}
\end{figure}

To gain further insights into the transmission--distance behaviour, we compared these results to the state-of-the-art hydrodynamic simulation \textsc{THESAN} \citep{2022MNRAS.512.4909G,2022MNRAS.511.4005K,2022MNRAS.512.3243S}, which is a large-volume ($L_\mathrm{box} = 95.5~\mathrm{cMpc}$) radiation-magnetohydrodynamic simulation that simultaneously models galaxy formation and the radiative transport into the IGM. 
It is convenient for comparisons to convert the average transmission $\mathcal{T}(r)$ relation into the transmission {\it relative} to the mean transmission at that redshift.  Figure \ref{fig:comp_THESAN} shows this normalized Ly$\alpha$ transmission, $\mathcal{T}(r)/\bar{T}-1$, as a function of distance from nearby galaxies as obtained from our observations in the lower and intermediate redshifts regimes (right panel) and those seen in the \textsc{THESAN} simulations \citep[][private communication]{2022MNRAS.512.4909G}. 
The observed transmissions are re-normalized to the mean value ($\bar{T}_\mathrm{los}$) along our specific sightline within each redshift range under consideration -- $\bar{T}_\mathrm{los}=0.0803$ for $5.3< z< 7.0$ and $\bar{T}_\mathrm{los} = 0.0023$ for $5.7< z < 6.14$.  
The analysis of the \textsc{THESAN} simulation computes the transmission around galaxies of $M_\ast>10^9~M_\odot$ at different redshifts $z=5.5\textrm{--}6.9$ in the (fiducial) \textsc{THESAN}-1 simulations.  The mean transmission value, $\bar{T}$, at each redshift is denoted in the figure legend.

The monotonic increase of transmission with $r$, from the very low value at the galaxies, up to the cosmic average transmission level, that we observed at $z=5.3\textrm{--}5.7$, is similar to the predicted behavior at the same redshift $z\sim5.5\textrm{--}5.7$ in \textsc{THESAN}, where the overall mean transmission is also comparable to the observed value.

Interestingly, the \textsc{THESAN} simulations at higher redshifts do show a qualitatively similar peaked transmission curve to that seen in the observational data that we have presented in this paper.  Quantitatively there are differences: the peak first really becomes evident in the \textsc{THESAN} simulation at $z=6.6$, as opposed to the $z \sim 5.9$ of our observations.  It is quite noticable however that the mean transmission along our line of sight at $z \sim 5.9$ is considerably lower than the \textsc{THESAN} $\bar{T}$ at $z=6.0$, and actually rather similar to that at $z=6.6$, implying that the reionization history along this line of sight differs from the average but within the observed spatial variations \citep{2018MNRAS.479.1055B,2018ApJ...864...53E}.  In other words, the \textsc{THESAN} curve and our own appear to be much more similar when compared at the same mean transmission value $\bar{T}$ than at the same redshift.
Having said that, there are other differences. Not least, the peak shows up at somewhat smaller $r$ in \textsc{THESAN} and the behavior at much larger $r$ also look different.

Regardless of some differences, this general qualitative similarity of our observational transmission curve with the results of these state-of-the-art simulations should be taken to strengthen our conclusion that the ionizing radiation determining the Ly$\alpha$ transmission of the IGM is dominated by the cosmic ionizing background at low redshift ($z\sim5.5$), but, at least along this line of sight, by the local ionizing effects of nearby galaxies at $z\sim5.9$.

We stress that all the results presented here are based on only a single line of sight.  Indeed, the key features in Figure \ref{fig:Tavg}, i.e., the peaked shape of $\mathcal{T}(r)$ that is seen at $z\sim5.9$ are coming from a small number of galaxy overdensities, of transmission spikes, and the association between the two.  It should be emphasized, however, that these data represent just the first 16\% of the EIGER survey, with five additional sightlines to be observed.  Their combined measurements will improve these constraints and inform new generations of models.
These will also enable us to examine the nature of galaxies owing to the local ionizing radiation (e.g., mass dependence of the escape fraction), which will be carried out in future papers.

\section{Summary}
\label{sec:summary}

In this paper, we introduced the EIGER survey, a large \textit{JWST}/NIRCam WFSS campaign that has been designed to produce large samples of {\Oiii}-emitting galaxies at $5.33 < z < 6.93$ in the fields of quasars for which we have very deep high resolution spectroscopy.  We have presented our first set of 117 {\Oiii}-emitting galaxies along the line of sight towards the ultraluminous $z=6.327$ quasar, J0100$+$2802.

The quasar resides in a strong overdensity, traced by 24 galaxies, and two other prominent overdensities also exist, one within the quasar near zone ($z=6.18$) and the other in the background ($z=6.78$).

One or more galaxies were identified within 200~pkpc and 105~$\mathrm{km~s^{-1}}$ of four metal-absorption systems, either as their physical host galaxies or as neighboring gas-galaxy associations embedded in a common larger-scale structures.

As the main focus of this paper, we carried out a cross-correlation analysis between the galaxies and the IGM transmission of Ly$\alpha$ and Ly$\beta$ in three redshift regimes and determined the average transmission as a function of distance from galaxies, $\mathcal{T}(r)$.

At the lowest redshifts $5.3 < z < 5.7$, $\mathcal{T}(r)$ increases monotonically with radius away from galaxies.  This effect is observed at lower redshifts and reflects a higher density of gas in the immediate vicinity of galaxies and the effect of a more or less uniform cosmic ionizing background, similar to observed at $z\sim2$--3.

At higher redshifts, $5.7 < z < 6.14$ (where the IGM is shielded from the quasar's ionizing radiation by a DLA) we see a different behaviour.  There are clear associations between galaxies and transmission spikes with separations of $\sim 5~\mathrm{cMpc}$.   The $\mathcal{T}(r)$ curve initially rises away from galaxies, as before, but sharply peaks at a distance of $\sim 5~\mathrm{cMpc}$, beyond which it rapidly drops again. 
We interpret this peak as indicating that the IGM transmission at this redshift is being produced by ``local'' galactic sources of ionizing radiation which dominate, at 5~cMpc, over the much weakened cosmic ionizing background.

Finally, as we enter the near zone regime around the quasar, at $6.15 < z < 6.26$, where the extra ionizing radiation from the quasar overwhelms other ionizing sources, the $\mathcal{T}(r)$ curve reverts to the monotonic behaviour seen in the lowest redshift range.

Overall, these results represent convincing evidence that local ionizing radiation from galaxies is responsible for producing the last traces of transmission of the IGM at $z \sim 5.9$, at least along this line of sight.

These early results demonstrate the unprecedented power of \textit{JWST} and particularly the WFSS observing mode of NIRCam to produce impressively large and complete samples of {\Oiii}-emitting galaxies at $5.3 < z < 7.0$.  Companion papers will explore other science topics that are enabled by this data set.

Our EIGER survey will conduct almost identical observations in five other fields, all centered on luminous quasars in the EoR.  The full survey, to be completed in the first year of \textit{JWST} science operations, should enable us to elucidate  the processes of cosmic reionization and early galaxy assembly.

\facilities{JWST (NIRCam)}

\software{
Python,
numpy \citep{harris2020array},
scipy \citep{2020SciPy-NMeth},
Astropy \citep{2013A&A...558A..33A},
Matplotlib \citep{4160265},
MIRaGe \citep{2022ascl.soft03008H},
PypeIt \citep{2019ascl.soft11004P}
}

\begin{acknowledgments}

We thank Enrico Garaldi, Rahul Kannan, Aaron Smith and the \textsc{THESAN} team for providing us with their simulation data, and Norbert Pirzkal for helpful advice on the NIRCam observing program.

This work is based on observations made with the NASA/ESA/CSA James Webb Space Telescope.  The data were obtained from the Mikulski Archive for Space Telescopes at the Space Telescope Science Institute, which is operated by the Associations of Universities for Research in Astronomy, Inc., under NASA contract NAS 5-03127 for \textit{JWST}.  These observations were taken under GTO program \# 1243.   

The participation of SL as an Interdisciplinary Scientist in the JWST Flight Science Working Group 2002-2022 has been supported by the European Space Agency.  His earlier involvement in the development of the JWST 1996-2001 was supported by the Canadian Space Agency.

Some of the data used in this work were obtained at the European Southern Observatory under ESO programme 096.A-0095, with the 6.5 meter Magellan Telescopes located at Las Campanas Observatory, Chile, and at the W. M. Keck Observatory, which is operated as a scientific partnership among the California Institute of Technology, the University of California, and the National Aeronautics and Space Administration.

This work has been supported by JSPS KAKENHI Grant Number JP21K13956 (DK).

\end{acknowledgments}


\bibliography{ads}
\bibliographystyle{aasjournal}



\appendix

\end{document}